\documentclass[twocolumn,aps,pre,superscriptaddress,floatfix]{revtex4-1}
\usepackage{graphicx}
\usepackage{epsfig}
\usepackage{thumbpdf}
\usepackage{amsfonts}
\usepackage{times}     
\usepackage{float}
\usepackage{indentfirst}   %
\usepackage{amsmath}
\usepackage{epstopdf}
\usepackage{textcomp}
\usepackage{tocvsec2}
\usepackage{epsfig}
\usepackage{xcolor}
\usepackage{comment}
\usepackage{CJK} 

\usepackage[section]{placeins}
\makeatletter\AtBeginDocument{%
     \expandafter\renewcommand\expandafter\subsection\expandafter
       {\expandafter\@fb@subsecFB\subsection}%
     \newcommand\@fb@subsecFB{\FloatBarrier
     \gdef\@fb@afterHHook{\@fb@topbarrier \gdef\@fb@afterHHook{}}}
     \g@addto@macro\@afterheading{\@fb@afterHHook}
     \gdef\@fb@afterHHook{}
  }
\usepackage[colorlinks=true,linkcolor=black]{hyperref} 

\newcommand{\be}{\begin{equation}}
\newcommand{\ee}{\end{equation}}

\begin{document}
\begin{CJK}{UTF8}{gbsn}

\title{Small world yields optimal public goods in presence of both altruistic and selfish cooperators}

\author{Pengbi Cui (崔鹏碧)}
\affiliation{Web Sciences Center, University of Electronic Science and Technology of China, Chengdu 611731, China}
\affiliation{Big Data Research Center, Univesrsity of Electronic Science and Technology of China, Chengdu 611731, China}

\author{Zhi-Xi Wu}\email{wuzhx@lzu.edu.cn}
\affiliation{Institute of Computational Physics and Complex Systems, Lanzhou University, Lanzhou Gansu 730000, China and Key Laboratory for Magnetism and Magnetic Materials of the Ministry of Education, Lanzhou University, Lanzhou Gansu 730000, China}

\author{Tao Zhou}\email{zhutou@ustc.edu}
\affiliation{Web Sciences Center, University of Electronic Science and Technology of China, Chengdu 611731, China}
\affiliation{Big Data Research Center, Univesrsity of Electronic Science and Technology of China, Chengdu 611731, China}

\begin{abstract}
Empirical studies have shown that individuals' behaviors are largely influenced by social conformity, including punishment. However, a coevolutionary theoretical framework that takes into account effects of conformity on individuals' punishment behaviors has not been put forward yet. Herein we propose a coevolutionary game model to extend the theory of cooperation with conformity in spatial public goods game by considering pool punishment, as well as two converse feedback modes of conformity that strongly affect cooperators' punishment behaviors. We focus on how different parameters and spatial structures govern evolutionary dynamics on three different kinds of networks by employing mean-field analysis based on replicator dynamics and Monte Carlo simulations. On regular lattices, defectors are overall extincted since cooperators, especially selfish cooperators, have great evolutionary advantages due to strong network reciprocity, and at the same time the number of altruistic cooperators decays. Conversely, abundant shortcuts in regular random networks lead to the prevalence of altruistic cooperators, but cooperators suffer from free-riding behaviors of defectors. Of particular interest, we find that small-world topology can simultaneously help cooperators successfully outperform defectors by means of strong network reciprocity, and enable rich contacting opportunities with defectors to facilitate the expansion of altruistic cooperators. Therefore, we clarify that small world is the optimal topology subject to the dominance of altruistic cooperators.
\end{abstract}

\maketitle
\end{CJK}

\section{Introduction}
\label{sec:introduction}
The first-order social dilemma in evolutionary game theory is that the well-being of the population depends only on the level of cooperation while defection is the best choice for individuals. Kin selection~\cite{foster2006}, reputation~\cite{alexander1987}, reciprocity~\cite{axelrod1984} and group selection~\cite{nowak2006} are well-known mechanisms to promote cooperation. Besides, punishment is widely accepted as an useful tool to repel defection and to facilitate cooperation~\cite{fehr2000,fehr2002,gachter2008,herrmann2008}. At the same time, the second-order social dilemma resulting from the fact that punishers have to bear extra substantial punishment cost remains a trouble, because this can weaken punishers' persistent monitoring ability and sanctions on wrong-doers~\cite{panchanathan2004,sigmund2010,perc2012}. Accordingly, some researchers tried to seek more efficient strategies or mechanisms to address the above issue~\cite{hauert2002,mathew2009,boyd2010,cui2014,chen2014}. 

Recently some experiments suggest that humans prefer pool punishment over peer punishment for maintaining the commons~\cite{traulsen2012}. Also, pool punishment is widely exploited in reality to mitigate the free riders' destructive
potential. The cost of pool punishment is shared by everyone, which thus reduces both financial burdens and risk of being revenged~\cite{nikiforakis2008}. Some third organizations, such as modern courts and police systems, could timely identify and punish defectors and thus can be considered as implement institutions of the pool punishment. Therefore, the problems about antisocial punishment~\cite{herrmann2008} and retaliation~\cite{nikiforakis2008} could be to some extent solved. Taken together, that is why pool punishment is an important symbol of modern civilized society, and attracts much attentions recently. However, maintaining costly pool punishment may still result in the second-order dilemma and thus erode cooperators. More realistic and powerful mechanisms are still needed.

On the other hand, substantial empirical studies proceeded by economists, psychologists and sociologists have shown that people tend to conform to other group members or the majority of their community. In individual psychology problems~\cite{insko1985,campbell1989}, voting situations~\cite{coleman2004,yang2013}, evolutionary game theory~\cite{cui2013,szolnoki2016} and so on, such phenomenon is named as majority rule, peer influence or social impact. Specifically, Hilbe et al. uncovered a close relationship between pool punishment and majority vote in terms of the prevalence of cooperation~\cite{hilbe2014}. Cui and Wu~\cite{cui2013} analyzed a variant of the prisoner's dilemma game (PDG) where an individual's strategies are affected by others' and showed the positive role of conformity in the emergence of large-scale cooperation. Moreover, some laboratory experiments on social dilemmas have pointed out that, under certain conditions, social impact involves not only human behaviors but also the punishment on these behaviors~\cite{horne2016}. We also note that Ref.~\cite{brown2000} indicates two different response modes of individuals to the influence exerted by conformity: positive feedback (the majority rule) and negative feedback (the minority rule). Therefore, understanding how punishment driven by different response modes of conformity sustains public goods in evolution systems of plentiful individuals is imperative. The model considering both pool punishment and conformity on punishment behaviors of individuals may be a little complex, but indispensable and more realistic. However, to the best of our knowledge, little attentions has been paid to this subject. 

In this paper, we focus on the public good game (PGG) which provides a good theoretical framework to investigate the evolution of cooperation in presence of both pool punishment and conformity. The reason to adopt PGG is twofold. Firstly, PGG involves a goup of players, which is suitable to test the effects of pool punishment. Secondly, PGG is widely accepted as the closest model to mimic money-seeking organizations such as banks, profit funds or listed companies: attracting capital and sharing investment gains together.   

This paper focuses on how punishment behaviors driven by peer influence govern the evolution of cooperation on different networks. In particular, we classify cooperators into two groups: altruistic cooperators (ACs) and selfish cooperators (SCs) according to two different feedback modes under peer influences. Our aim is to identify the optimal conditions including both parameter values and topological features for overall cooperators and the prevalence of altruistic cooperators. In detail, we find that regular lattice (RL) is a perfect breeding ground for selfish cooperators, where altruistic cooperators and selfish cooperators exclude each other. Instead, on a regular random network (RRN) the two types of cooperators behave as a mutual assistance alliance to resist defectors which are in the dominated position in most parameter regions. Finally, the combination of clustering and shortcuts, say the small-world network, is demonstrated to be the optimal topology to sustain altruistic cooperation and public goods under suitable parameter conditions. In addition, both simulations and analysis strongly support our proposed physical interpretation of the emergence of abundant altruistic cooperative behaviors on the small-world networks.

This paper is organized as follows. We firstly give a detailed description of our model on Sec.~\ref{sec:model}. In Sec.~\ref{sec:results}, in turn, we fully explore the model on regular lattices (RLs) in~\ref{subsec:rls}, random regular networks (RRNs) in~\ref{subsec:rrns} and small-world networks based on 2D RLs in~\ref{subsec:smworld}; respectively. Both agent-based Monte Carlo simulation (MCS) and theoretical analysis based on mean-filed theory are employed, which are in good accordance with each other. Finally, we conclude this paper in Sec.~\ref{sec:discuss}.

\section{Model}
\label{sec:model}
On the network of size $N$, an over-lapping game group contains all the nearest neighbors of the focal individual besides itself, where each group member simultaneously plays the public goods game (PGG). At the same time, each individual $i$ participates in $k_{i}$ games initiated by her neighbors, in addition to holding a public goods game (PGG) with size $G_{i}=k_{i}+1$ (together with all his neighbors); where $k_{i}$ is the number of focal individual's connections with others (i. e., the degree). Therefore, each individual $i$ simultaneously plays $k_{i}+1$ PGGs by holding the same strategy. In accordance to the definition of PGG, each cooperator makes a contribution of $1$ to the public good, while defectors (Ds) contribute nothing. Subsequently, the sum of all the contributions in a group is multiplied by the synergy factor $r>1$, which takes into account synergistic effects of cooperation. The resulting amount is then equally shared among the members of the group. In detail, there will be two different cases for $i$ if $i$ is a cooperator: (1) If $i$ carries out punishment at a probability $p$ in an interaction, its payoff would be $\Pi_{P}=rn_{C}/G_{i}-1-n_{D}\alpha /n_{P}$; (2) otherwise the payoff of $i$ is $\Pi_{C}=rn_{C}/G_{i}-1$. $n_{P}$ ($n_{C}$) is the number of punishers (cooperators) in the group. $\alpha$ represents the punishment fine that each defector incurs in presence of punishers. In the case that $i$ is a defector, $\Pi_{D}=rn_{C}/G_{i}$ if $n_{P}=0$, or else $\Pi_{D}=rn_{C}/G_{i}-\alpha$.

This paper mainly investigates interactions between evolution of cooperation and spatial structures of networks or parameter conditions. In detail, there are two types of cooperators: altruistic cooperators (ACs) and selfish cooperators (SCs) which would carry out a punishment on defectors in the same group at a different probability function $p$. $p$ is the feedback function of the number of cooperators (defectors) $n_{C}$ ($n_{D}$) in one group, showing two opposite forms according to conformity preferences of the cooperators:
\begin{equation}
\begin{cases}
p_{a}=A\frac{n_{D}}{G} \quad \text{AC}\\
p_{s}=A\frac{n_{C}}{G} \quad \text{SC}
\end{cases}
\label{eq:asfunztion1}
\end{equation}
The two response modes (i.e., conformity preference) quantitatively assumed here have been mentioned and discussed by Ref.~\cite{brown2000}, showing its rationality. It can be found that $p_{a}  =  A-p_{s}$, and the size of the game group is $n=k+1=n_{C}+n_{D}$. Parameter $A \in [0,~1]$ quantifies the sensitivity of feedback, larger $A$ indicates larger differences between AC and SC, or more sensitive response of ACs (SCs) to the number of defectors (cooperators) in the group. Relating to the reality, this feedback mechanism defined by Eq.~\ref{eq:asfunztion1} reveals that ACs incline to step forward to punish rather than stand aside, regardless of the considerable cost of punishment caused by increasing number of defectors. In contrast, SCs just take actions when there are more cooperators (thus less defectors) to sharing the punishment cost, in purpose of reserving their payoffs firstly. This is the reason that herein the two types of cooperators are respectively named as altruistic cooperators and selfish cooperators. Actually, ACs and SCs construct a new kind of social dilemma other than the traditional dilemma consisting of cooperators and defectors. As 'prudent' guys, SCs could be considered as second-order free riders because they preserve higher payoffs than those ACs that abhor evil as a deadly foe by doing less to fight against wrong-doers. For simplicity of model, it must be stressed that in our model this feedback mechanism is only stated on punishment behaviors of cooperators, which means that no other mechanisms or incentives like what mentioned in Refs.~\cite{cui2013,cui2014} are introduced to drive defectors to punish others, or change strategies of individuals. 

\begin{figure}[h]
\includegraphics[width=0.5\textwidth]{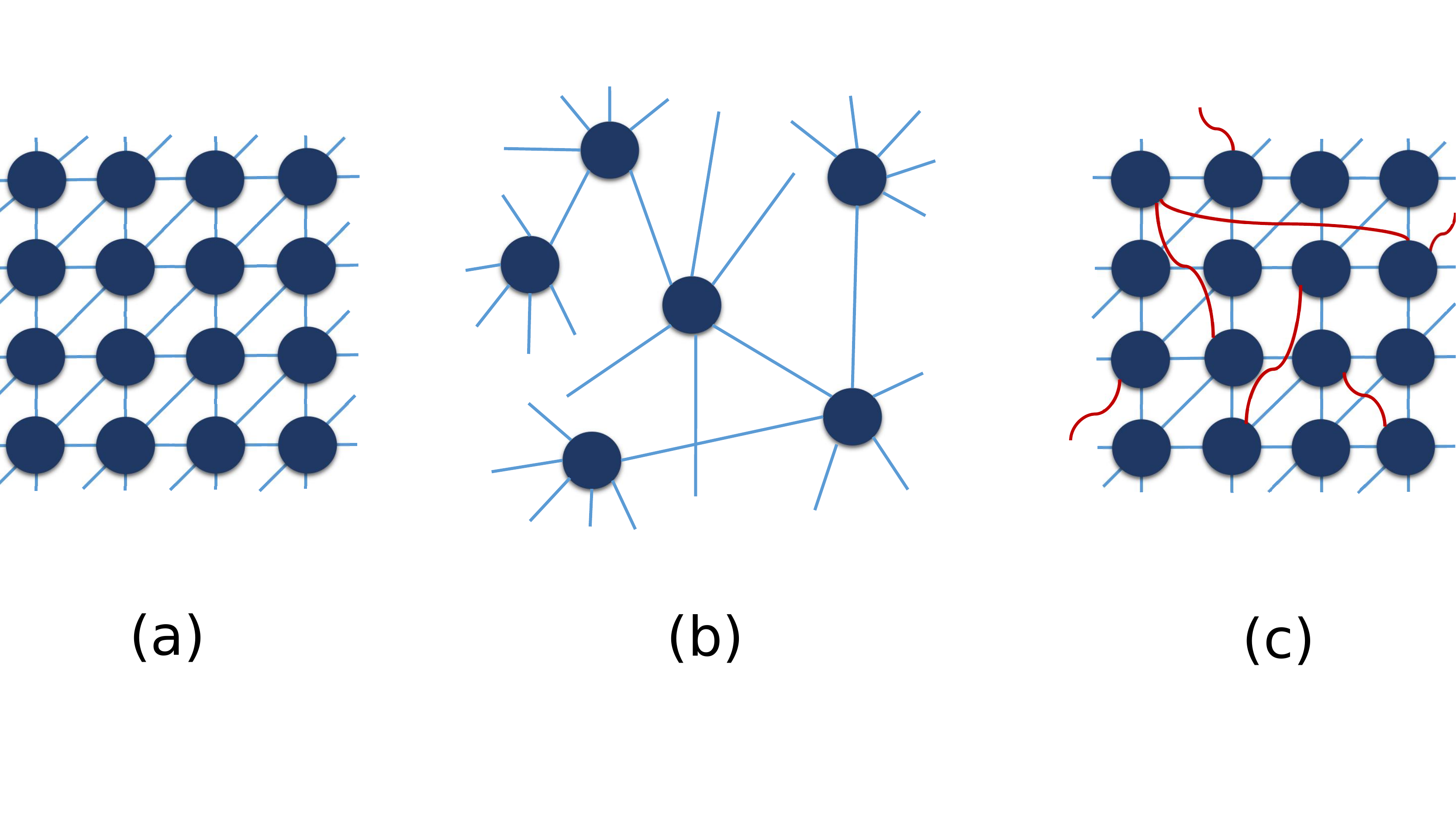}
\vspace{-1cm}
\caption{Illustrations of the three networks to be employed for numerical treatment. In detail, (a) Regular lattice networks (RLs) i.e., Hexagonal lattice; (b) Regular random networks (RRNs) in which connections are randomly distributed among nodes; (c) small-world networks generated by our developed algorithm, which contains both local connections (dark blue edges) and patterns of shortcuts (red edges). In the three networks, the degree of any particular node remains constant i.e., $k=6$. Also, both self-connections and multiple connections are avoided.
}
\label{fig:threenetwork}
\end{figure}

Fig.~\ref{fig:threenetwork} presents three different networks for detailed numerical treatment in this study, as well as a brief description about these networks. Especially, we develop a new algorithm which is derived from the original small-world models~\cite{watts1998,newman1999} to generate small-world networks on an original 2D RL for sake of theoretical analysis. According to the algorithm, we generate the networks by starting with a regular 2-dimensional ($d=2$) lattice of size $N=L^{d}$ to guarantee dense connections. Then each of the connections in the regular lattice is in turn, independently and randomly exchanged one of its ends with that of another randomly-selected connection with probability $p$ to produce patterns of shortcuts.

In numerical treatment, MCS is employed to update the strategies and conformity preferences of players. And random sequential updates are adopted to control the evolution of the population. Initially each player fixed on the networks is randomly and independently designated as cooperators or defectors, and altruistic kind or selfish kind. Each simulation procedure contains $N$ times of the following steps such that every one owns one and only one chance to change its strategy and conformity preference on average: (1) A randomly selected player $i$ accumulates its payoff $\Pi_{i}$ by playing $G_{i}$ PGGs with its $k$ nearest neighbors as one member of the $G_{i}$ groups. The randomly-chosen nearest neighbors $j$ also obtains its payoff $\Pi_{j}$ in the same way. (2) Then $i$ simulates both the strategy and conformity preference of $j$ with probability $W_{j\longleftarrow i}=1/[1+\exp((\Pi_{i}-\Pi_{j})/\kappa)]$. The fermi study function implies players owning higher payoffs are advantaged, while adoption of those of a player performing worse is still possible. $\kappa$ curves the noise of the uncertainty in the adoption. Without loss of generality we set $\kappa=0.1$ throughout this paper. The simulations are performed until the system has reached steady state, i. e., the numbers of ACs, SCs and Ds keep stable.

The final densities of all four strategies ($\rho_{s}$) are obtained after at least $1.0\times 10^{4}$ Monte Carlo steps (MCS) to guarantee equilibrium existence, and averaged over $20-50$ independent realizations to insure a low variability. The size of RRNs is $N=40000$, and the size of RLs or the small-world networks based on 2D RLs is $N=200\times 200$. Throughout this paper, degree of each node in employed networks is $k=6$.

\section{Results}
\label{sec:results}
\subsection{The results on regular lattice networks}
\label{subsec:rls}
In this subsection, we focus on how punishment fine, feedback sensitivity, synergy factor of PGG govern the evolution of cooperation on RLs. We simulate the evolutionary dynamics ruled by our game model as defined in Sec.~\ref{sec:model}, as well as a theoretical analysis listed  detailedly in Appendix.~A.

\begin{figure}[h]
\includegraphics[width=0.5\textwidth]{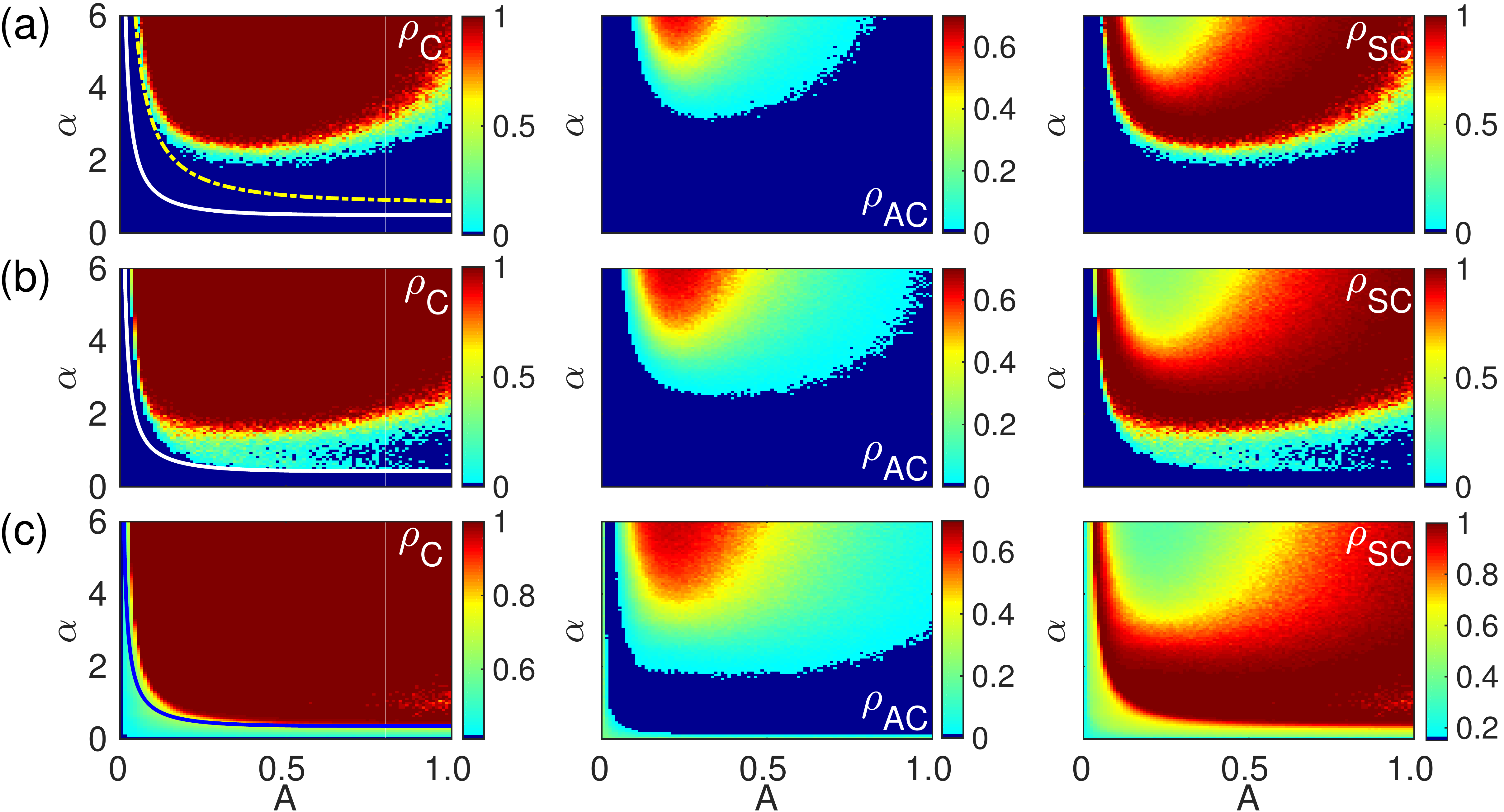}
\caption{Fractions of cooperators $\rho_{C}$, altruistic cooperators $\rho_{AC}$ and selfish cooperators $\rho_{SC}$, plotted versus $A$ and $\alpha$ on Hexagonal lattice. The values of the synergy factor are $r=3.5$ (a), $r=4.0$ (b), and $r=4.5$ (c); respectively. The solid lines represent the analytical predicted critical boundaries, while the dashdotted lines the semi-analytical prediction. And the estimated value of $f_{1}$ in semi analysis is $0.75$ for $r=3.5$. To differentiate, we let $f_{1}$ ($f_{2}$) denote density of ACs (SCs) in theoretical or semi  analysis, and see Appendix.~A for more information of $f_{1}$ ($f_{2}$)).}
\label{fig:hexAalpha}
\end{figure}
\begin{figure}[h]
\includegraphics[width=0.5\textwidth]{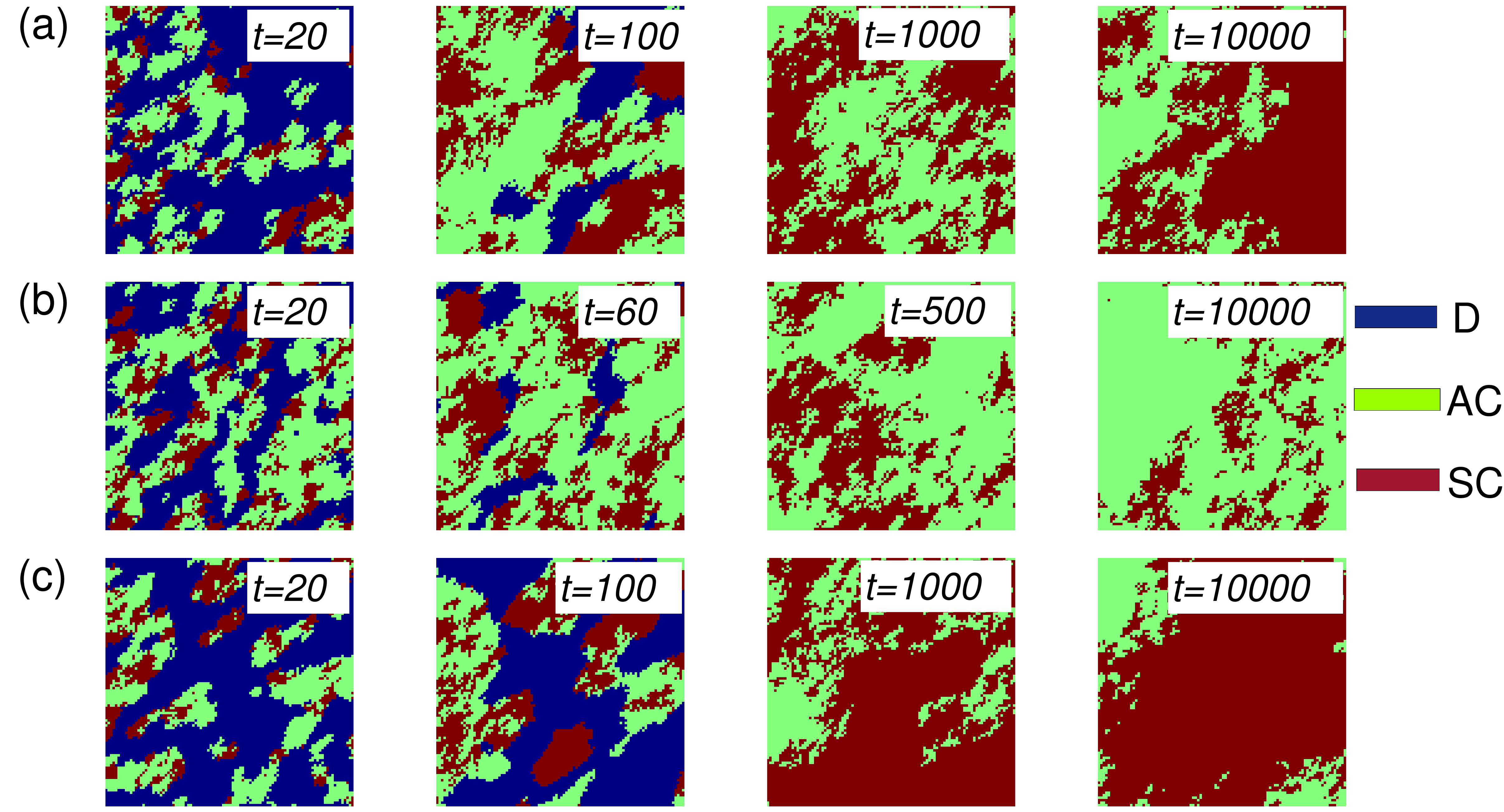}
\caption{Typical snapshots of the simulation grid for three different values of $A$: $A=0.13$ (a), $A=0.235$ (b) and $A=0.42$ (c) at different time, where $r=3.5$ and $\alpha=6.0$. Herein, $100\times 100$ windows of computer simulations are shown. Defectors are represented by blue, altruistic cooperators by green and selfish cooperators by red. The last two snapshots of each subfigures show a majority-like evolution phenomena: whoever is superior in numbers after defeating the defectors is dominator. Additionally, subfigure (b) depicts an optimal situation for ACs.}
\label{fig:spatialhexr35}
\end{figure}

Fig.~\ref{fig:hexAalpha} presents the dependence of fractions of cooperators $\rho_{C}$, altruistic cooperators $\rho_{AC}$ and selfish cooperators $\rho_{SC}$ on $A$ and $\alpha$ on Hexagonal lattice for different $r$. It is obvious that cooperators are able to persist and even prevail in most parameter region, which confirms positive roles of lattices with dense connections in promoting cooperators~\cite{szabo2007} especially selfish cooperators. On RLs, cooperators in spite of ACs and SCs could easily get together to form clusters to resist defectors, so that there are less defectors around cooperators or in a group containing cooperators. This is called network reciprocity~\cite{szabo2007,nowak2006,szolnoki2009}. A direct illustration of the evolution is shown in Fig.~\ref{fig:spatialhexr35}. Large clusters of cooperators are formed in RLs. SCs driven by proportions of cooperators in the group are thus more likely to carry out punishment to expand and capture defective domains. Longer boundaries between SCs and Ds are found in Fig.~\ref{fig:spatialhexr35}(a)(c). Moreover, dense distributions of local cooperators resulting from high local clustering could share more punishment cost together~\cite{szabo2007}. Especially, for large value of $A$, SCs are more dominant by taking advantage of the fight of altruistic cooperators against defectors. This is also further verified by illustrations in Fig.~\ref{fig:spatialhexr35}(c), in which SCs capture most inner regions of cooperator clusters as ACs is resisting Ds at boundaries of these clusters. Consequently, SCs prevail over ACs in the defector-existed period (i.e., early stage of evolution). After the population being fully occupied by cooperators, the evolutionary dynamics could be essentially mapped into an opinion tuning process, because every cooperator owning equal payoffs. Consequently, whoever is superior in numbers is dominator. Finally, SCs are at a dominated position.

At the same time, we note that there always exists a small optimal parameter region (near $A=0.235$, and punishment should be strong enough i.~e. $\alpha>4.0$) where ACs instead thrive. Furthermore, larger optimal parameter regions are observed for higher $r$, which indicates that evolutionary advantages of ACs could be more easily enhanced by higher synergy factor. A clear micro-level picture could explain this. Unresponsive ACs brought about by small $A$ would exert too few punishment on Ds within the group, so as to leave the expand opportunities to SCs (Fig.~\ref{fig:spatialhexr35}(a)). Conversely, large $A$ means that ACs would be too sensitive to punish too many defectors of different groups, which reduces payoffs of themselves too much, leading to being devoured by SCs (Fig.~\ref{fig:spatialhexr35}(c)). In any case, strong punishment (high $\alpha$) is a premise to guarantee possible prevalence of ACs. Moreover, it can be observed in Fig.~\ref{fig:hexAalpha} that the two types of cooperators exclude each other (the parameter regions of their prevalence are not overlapped), but not totally. Close coexistence of ACs and SCs seems more or less hard to achieve, attributing to the majority-like evolution rule~\cite{krapivsky2003} mentioned above. Fig.~\ref{fig:latticeround} in Appendix.~B also illustrates dynamical behaviors of the system by plotting evolution fractions of different kinds of cooperators. Diverse relaxation time for different punishment fine $\alpha$ and $A$ could be firstly observed. Especially, the fractions of selfish cooperators have the rule of first decreased then rose, which reveals that SCs are better than ACs in getting together to from clusters to face defectors.

Besides, there exist two relatively constant critical values of feedback sensitivity and punishment fine: $A_{c}$ and $\alpha_{c}$ beyond which the dominance of cooperators in the population is established. In other words, strong effectiveness of punishment~\cite{brandt2003} and enough sensitive conformity feedback of individuals are simultaneously needed to sustain the public goods. 

Herein mean-filed theory based on well-mixed assumption can be used to analyze the stability of the evolution of the system and to further predict the boundary between full-cooperation phase and full-defection phase. Whatever, Fig.~\ref{fig:hexAalpha} shows that our theoretical analysis is able to curve the core of the evolution through giving existence condition of interior equilibria point. Please see Appendix.~A for more information about theoretical analysis. 

Overall, the above results suggest that the punishment driven by conformity is surprisingly a useful tool to suppress defection strategy in a population with dense local connections. This has not been stressed or even mentioned by previous researches involving punishment mechanism in evolutionary game theory. However, it can be concluded that the first order social dilemma is restrained while the second order dilemma still exists in RLs. Since SCs dominate the population as the second-order free riders under most parameter conditions. Therefore we shift our focus from RLs to RRNs in the following subsection.

\subsection{The results on regular random networks}
\label{subsec:rrns}
\begin{figure}[ht!]
\includegraphics[width=0.5\textwidth]{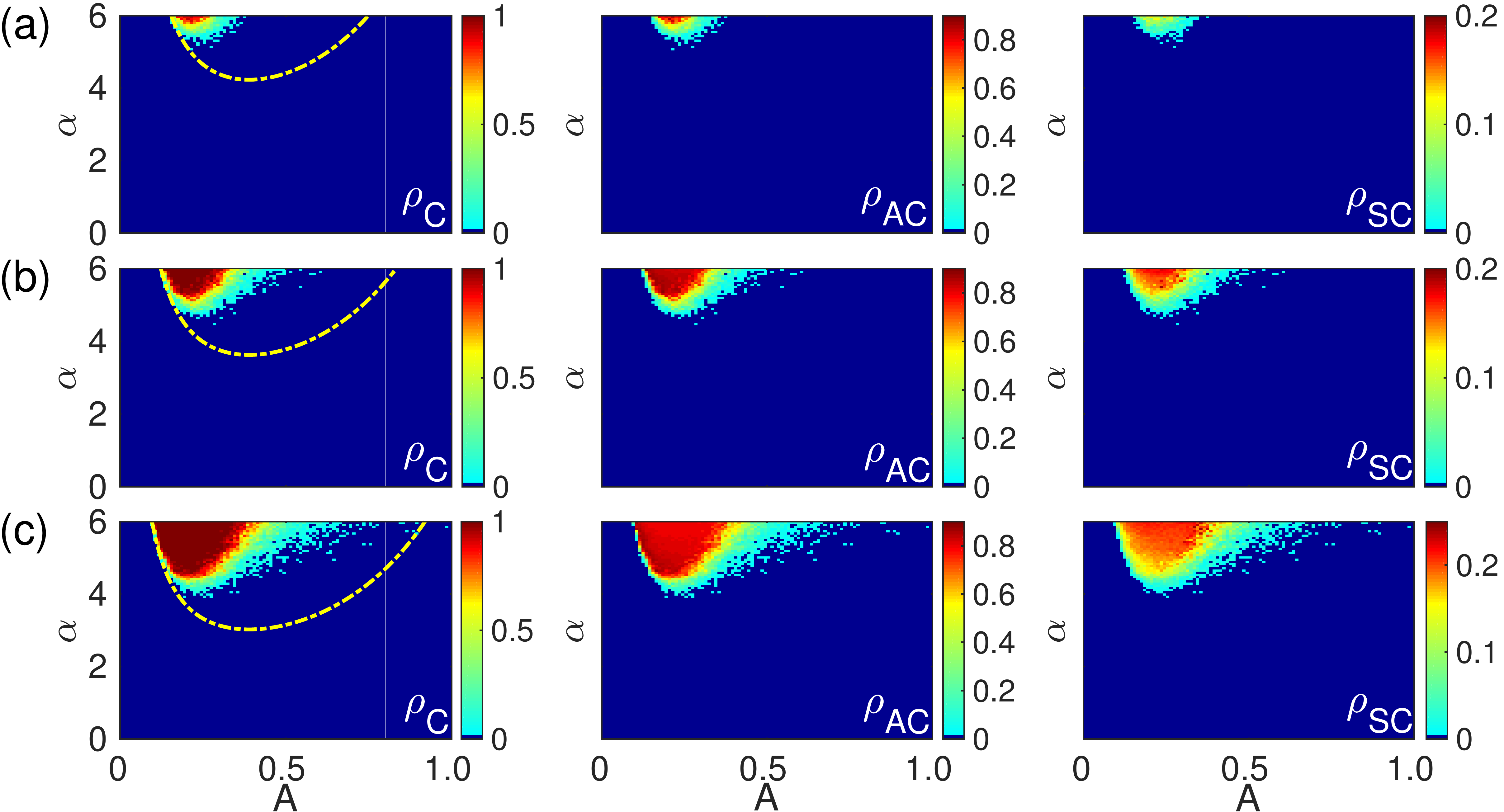}
\caption{Fractions of cooperators $\rho_{C}$, altruistic cooperators $\rho_{AC}$ and selfish cooperators $\rho_{SC}$, plotted versus $A$ and $\alpha$ on the RRNs. The values of synergy factor $r$ are: (a) $r=3.5$, (b) $r=4.0$ and (c) $r=4.5$, respectively. The dashdotted lines represent the semi-analytical predictions; where the estimated value of $f_{1}$ is: (a) $0.629$, (b) $0.618$ and (c) $0.621$, respectively.}
\label{fig:rraalpha}
\end{figure}
In comparison with the results of RLs, the results presented in Fig.~\ref{fig:rraalpha} give a totally different picture. Cs are suppressed to a certain extent while Ds hold evolution advantages in most parameter regions. Besides, as expected, higher returns of contribution (large $r$) extend the parameter regions of full-cooperation phase, but not too much. Less overlaps between PGG groups caused by sparse local clustering of RRNs is a crucial negative factor to cooperation. Because of this factor, network reciprocity between Cs is weak so that they fail to sustain competitive payoffs in comparison to Ds through frequently reciprocating with each other. Even worse, there are less punishers to share the cost of pool punishment together. This leads to a disappointing result that cooperators could not stop invasion of Ds in most parameter cases on RRNs.

\begin{figure}[ht!]
\includegraphics[width=0.5\textwidth]{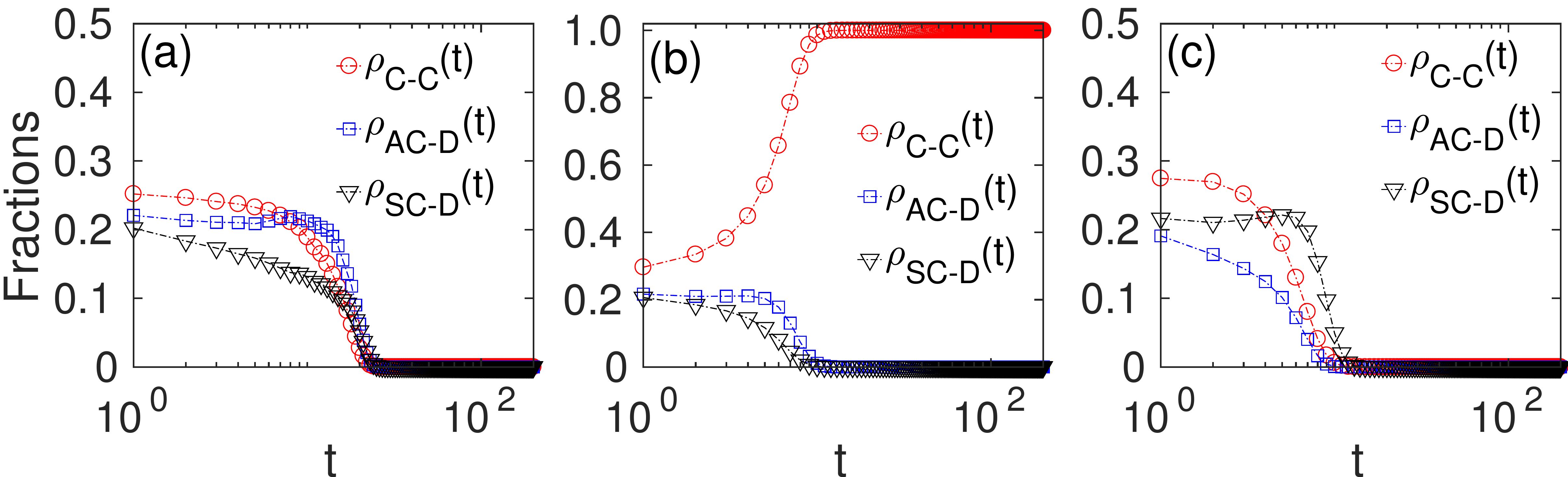}
\caption{Evolution of the fractions of three different strategy pairs on RRNs. The feedback sensitivity of the individuals is $A=0.09$ (a) $A=0.20$ (b) and $A=0.55$ (c), respectively. The other parameters are taken as $r=4.5$ and $\alpha=6.0$.}
\label{fig:fractioncd}
\end{figure}
\begin{figure}[ht!]
\includegraphics[width=0.5\textwidth]{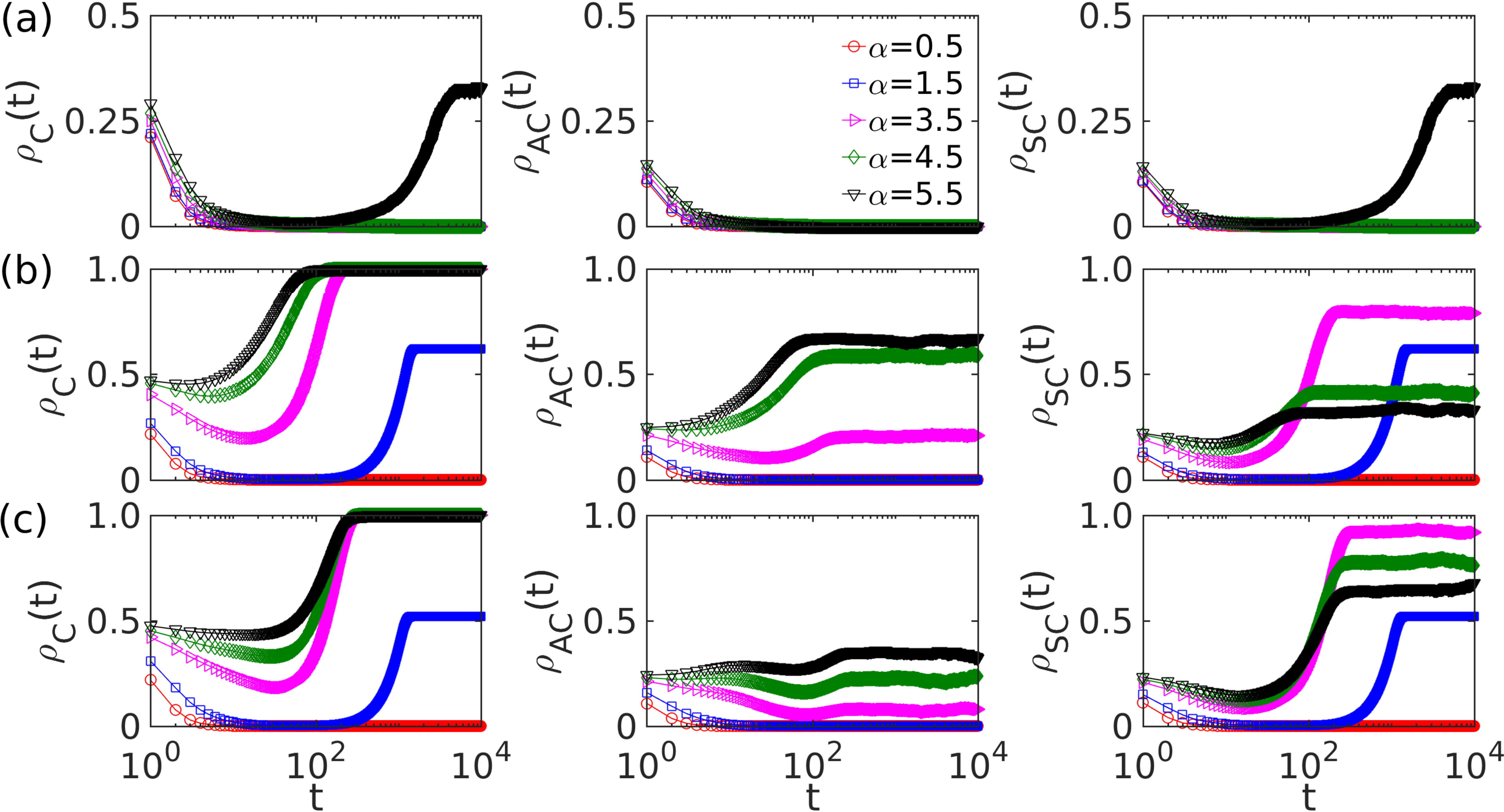}
\caption{Evolution of fractions of three different types of cooperators for different punishment fines on RLs. The values of feedback sensitivity are $A=0.03$ (a) $A=0.235$ (b) and $A=0.5$ (c); respectively. The other parameter is $r=4.0$.}
\label{fig:latticeround}
\end{figure}

However, it is surprising that ACs are able to surpass both Ds and SCs if two conditions strong punishment (large value of $\alpha$) and appropriate feedback sensitivity ($A\approx 0.2$) are simultaneously satisfied. Fig.~\ref{fig:fractioncd} provides some hints on how to give the interpretations of this phenomenon by plotting the evolution of the fractions of three different strategy pairs. From the results for RLs we know that SCs are more dependent on local supports from other cooperators, i.e., they need more surrounding cooperative players to support them and meanwhile to drive them to sanction the surrounding Ds. Therefore their expanding ability is limited on RRNS with sparse local connections.  While things become complicated for ACs under such circumstance. Overall, more shortcuts generated from random reconnections enable ACs to get in ouch with large number of Ds to stimulate more punishment on Ds with more opportunity. In case of small $A$, more strategy pairs of $AC-D$ (see Fig.~\ref{fig:fractioncd}(a)) in the beginning of evolution mean that insensitive or even unresponsive ACs would have high payoffs to confront with Ds; but they are finally prohibited or even absorbed by Ds like SCs due to lack of punishment to suppress these Ds. In contrast, the cost of punishment that excessively sensitive ACs curved by large $A$ should share may be too high to greatly reduce their payoffs, because too many Ds are sanctioned. Less strategy pairs of $AC-D$ are thus observed in Fig.~\ref{fig:fractioncd}(c). This explains why response sensitivity of ACs should be median, as shown in Fig.~\ref{fig:fractioncd}(b), under the parameter condition they could not only defeat Ds through enough and strong punishment (large $\alpha$) but also maintain higher payoffs than Ds at the borders of C clusters to finally prevent themselves from Ds' invasion, or even absorb them into C clusters. The evolutionary dynamics corresponding to Fig.~\ref{fig:fractioncd} are also supplemented in Fig.~\ref{fig:rrnround} of Appendix.~B, by giving evolution of different types of cooperators. Unlike what we have observed in Fig.~\ref{fig:latticeround}, the systems on RRNs can quickly reach a stable state due to lack of cooperators' clustering process, regardless of punishment fine and feedback sensitivity of the individuals.

Differing the mutually exclusion between ACs and SCs on RLs, Fig.~\ref{fig:rraalpha} shows that the parameter regions of positive fractions of ACs and SCs are highly overlapped; implying a strong reciprocity relationship like mutualistic symbiosis. It suggests that SCs, considered as the second-order free riders, should be only under the protection of ACs to survive. After all, SCs could sustained considerable payoffs through cooperating with ACs. 

Moreover, notice that both Fig.~\ref{fig:rraalpha} and Fig~\ref{fig:fractioncd} give two stable equilibrias: full cooperation and full defection. This means that majority-like rule still works after that all Ds have died out, which is independent of the network structures. Furthermore, the simulated phase regions are more or less in accordance with the predictions given by the semi-analysis through estimating value of $f_{1}$. It is strange, but easily understood. Large amount of shortcuts give rise to large contact areas between Cs and Ds. Hence large relatively closed clusters containing only Cs are not allowed to exist universally on RRNs~\cite{szabo2007}. The numbers of strategy pairs of $AC-D$ and $SC-D$ always close to each other at defector-existed stage of evolution (see Fig.~\ref{fig:rraalpha}), neither side have a big advantage in number. As a consequence, especially near the critical boundaries, strategy abundance behaves sharper fluctuations between full-C state and full-D state. We thus only have to estimate a constant fraction of ACs to enable a relatively precise analytical prediction (for more details see Appendix.~A). Fortunately, coincidence between simulated boundaries and dashdotted lines confirms the feasibility of our semi-analytical method.

On the whole, the second-order social dilemma are overcome to some extent on RRNs, but the problem of the first-order social dilemma becomes much more serious owing to absence of dense local connections. It seems hard to find an optimal topology on which not only the traditional social dilemma (the first-order social dilemma) but also the second-order social dilemma could be alleviated. Nevertheless, our detailed investigations on RLs and RRNs provide a crucial hint: dense local clustering is very favorable to prevalence of SCs while shortcuts largely facilitate the dominated roles of ACs in sanctioning or suppressing Ds. Hence the optimal topology must be a combination of dense local connections and considerable shortcuts. Naturally, small-world network is the first choice. Next we will proceed our study on this network, through both agent-based simulations and qualitative theoretical analysis. 

\subsection{The results on small-world networks based on 2D RLs}
\label{subsec:smworld}
In what follows, we mainly focus on how topology of small-world networks especially the number of shortcuts quantified by randomness $p$ affects the evolution of different strategies, under different parameter conditions. 

\begin{figure}[ht!]
\includegraphics[width=0.5\textwidth]{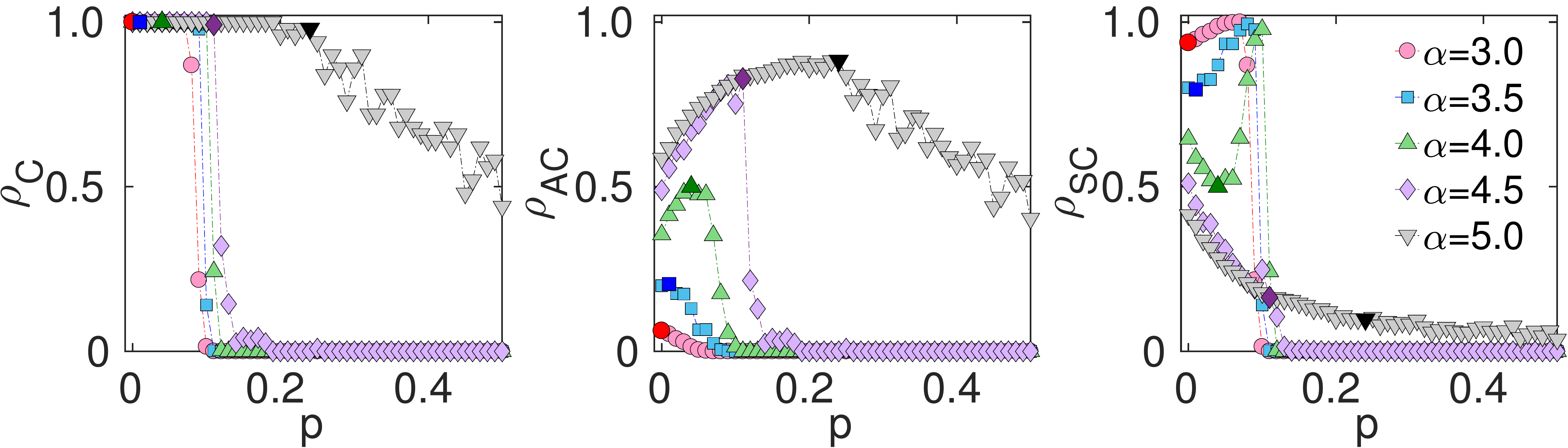}
\caption{Fractions of cooperators $\rho_{C}$, altruistic cooperators $\rho_{AC}$ and selfish cooperators $\rho_{SC}$, plotted versus randomness $p$ for the small-world networks based on 2D RLs. The results for different values of $\alpha$ are plotted. The other parameter are taken as $r=4.0$ and $A=0.235$. The dark marker for each $\alpha$ corresponds to the position of optimal randomness ($p^{*}$) where the population of ACs is maximum. We could also have a deep understand of the relationship between ACs and SCs by comparing the population size of ACs with that of SCs at this parameter point. 
}
\label{fig:samer40fp}
\end{figure}
We firstly plot the fractions of three populations: Cs, ACs and SCs in Fig.~\ref{fig:samer40fp} as function of network randomness $p$. Monotonous decay of cooperator populations with $p$ can be observed in the first panel of Fig.~\ref{fig:samer40fp}. The small-world networks based on 2D RLs approximates RLs as $p$ approaching to zero, and adversely RRNs if $p$ is large enough. Therefore, the arguments put forward in Subsec.~\ref{subsec:rls} and Subsec.~\ref{subsec:rrns} allow us to easily understand the changes of cooperator population with the randomness of networks. As we pointed out, dense local clustering resulting from small $p$ facilitates cooperators' evolution success, until the critical point of randomness that network reciprocity fails to sustain dominance of Cs. Furthermore, full-C phase could be easily achieved in regions of  strong network reciprocity which could be enlarged by punishment fine $\alpha$. 

Importantly and interestingly, we find in the middle panel of Fig.~\ref{fig:samer40fp} that in most cases with $p$ there exist optimal intermediate parameter regions centering on the peaks of ACs' fractions at $p^{*}$; where not only Cs rise to full dominance (capture all the population), but also ACs are evolutionary successful by successfully eliminating and outperforming others to form a stable coexistence with SCs as a result. This means that the small-world networks is the optimal topology we try to seek to alleviate both the first-order and the second-order social dilemma. More precisely, as shown in Fig.~\ref{fig:samer40fp}, larger punishment fines enable wider optimal regions; implying that only strong effectiveness of punishment guarantees the perfect performance of the small-world networks based on 2D RLs with more opportunity. Additionally, the third panel of Fig.~\ref{fig:samer40fp} presents a narrow optimal region of SCs when effectiveness of punishment is not strong enough, as $p$ is larger but still intermediate. Also, similar phenomena could be observed for other values of synergy factor such as $r=3.5$ and $r=4.5$ (please see Fig.~\ref{fig:2dws3545} listed in Appendix.~B). Microscopic mechanism behind these rich reported evolutionary outcomes is required to be given to make understanding of the evolutionary picture clear.

\begin{figure}[ht!]
\includegraphics[width=0.5\textwidth]{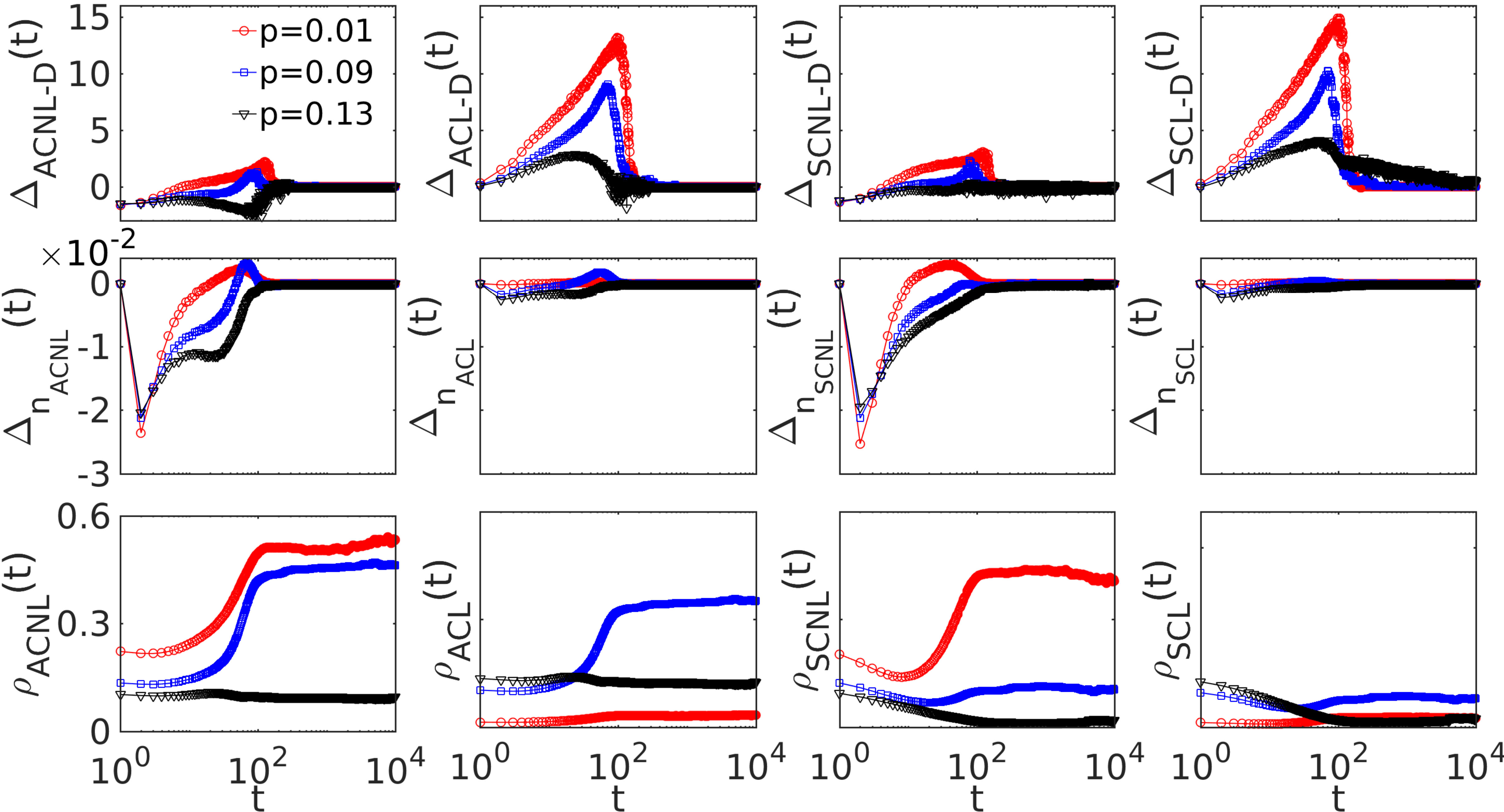}
\caption{Evolution of different statistics for three different values of randomness of the small-world networks base on 2D RLs. Different statistics are plotted to enable a deep interpretation of the results presented in Fig.~\ref{fig:samer40fp}. Top panels: the mean payoff gaps between different types of cooperators and their connected defectors. Middle panels: changing rate of edges $D-S$ to quantify how many edges change from $D-S$ to $S-S$ ($S=ACNL$, $ACL$, $SCNL$ and $SCL$) at each time step, which are normalized by the total number of edges in the networks. Bottom panels: evolution frequencies of four different types of cooperators. The other parameters are taken as $r=4.0$, $A=0.235$ and $\alpha=4.5$.
}
\label{fig:roundwsr40}
\end{figure}
\begin{figure*}
\includegraphics[width=0.9\textwidth, height=7.5cm]{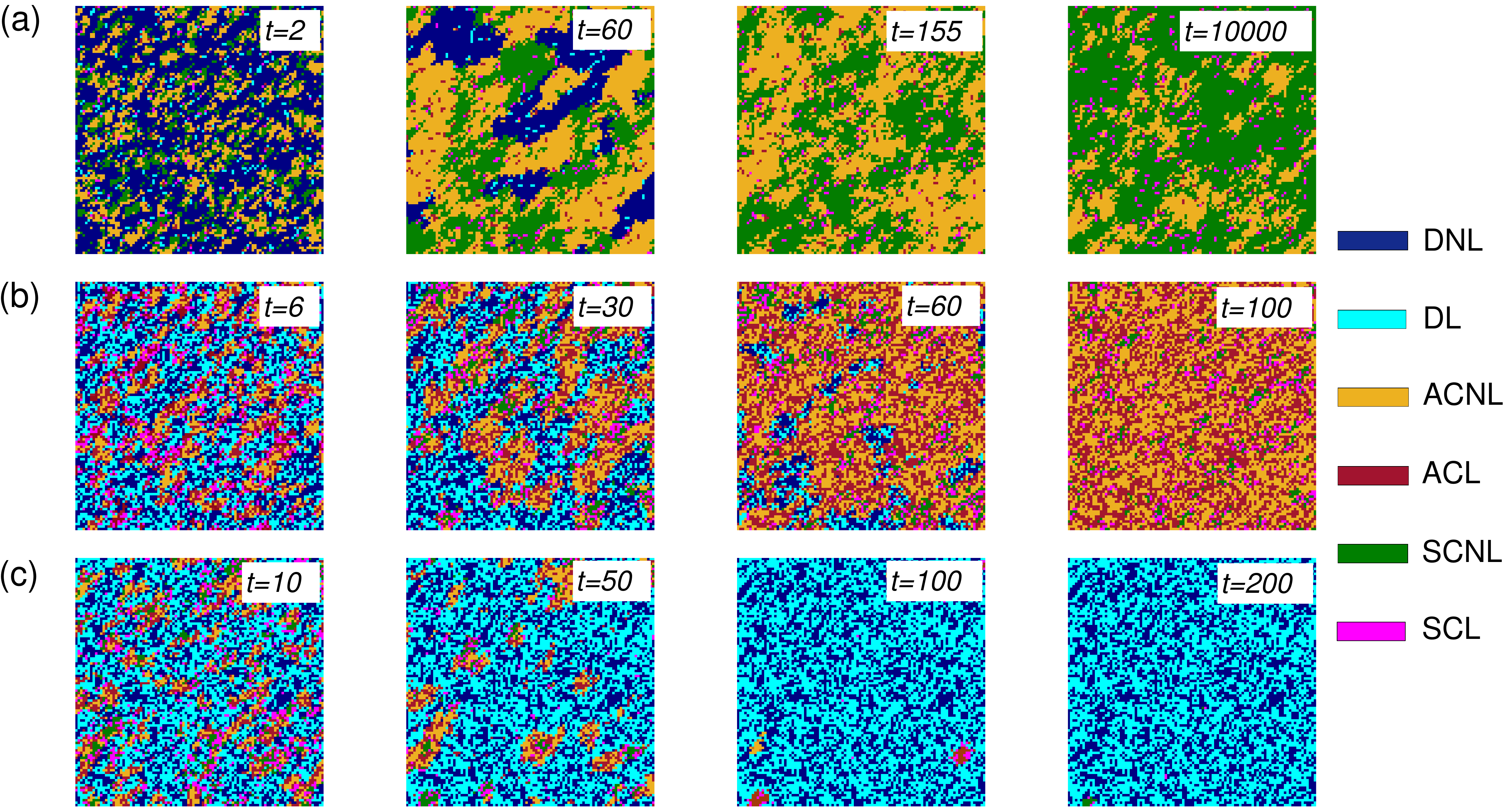}
\caption{Typical snapshots of the simulation grid for three different values of $p$: $p=0.01$ (a), $p=0.09$ (b) and $p=0.13$ (c) at different time. Here, $100\times 100$ windows of computer simulations are shown. For sake of interpretation of the evolution dynamics on the small-world networks, we further clarify the cooperators with the same conformity preference into two classes. In detail, defectors (altruistic cooperators, selfish cooperators) at the end of shortcuts are denoted as DL (ACL, SCL), while local defectors (altruistic cooperators, selfish cooperators) which are locally connected are denoted as DNL (ACNL, SCNL). This enables a more intuitive understanding of roles of shortcuts which containing long-range connections in the evolutionary dynamics. Moreover, the colors for corresponding strategies or types are labeled in above figure. The other parameters are taken as $A=0.235$, $\alpha=4.5$ and $r=4.0$.}
\label{fig:spatialws}
\end{figure*}
Subsec.~\ref{subsec:rls} and Subsec.~\ref{subsec:rrns} have listed detailed arguments to explain the evolution outcomes under two situations: $p=0$ and $p=1$. Now we center our attention on the mechanisms for behaviors of system at parameter region $0<p<1$. Essentially, effects of spatial topology of the small-world networks on changes in number of ACs and SCs can be explained by the comparison between networks reciprocity stemming from dense local clustering and odds of contacting with Ds afforded by shortcuts. Both Fig.~\ref{fig:roundwsr40} and Fig.~\ref{fig:spatialws} provide a more direct insight into the evolution behaviors of different strategies and conformity preferences. If $p$ is small, strong network reciprocity confer SCs and ACs enough payoffs to hold strategy (top panel of Fig.~\ref{fig:roundwsr40} show that both ACs and SCs own higher payoffs than their connected defectors do), while the number of contacted Ds of ACLs are less than they ought to be able to sanction. This implies that ACLs are not fully utilized in suppressing Ds. Positive peak of $\Delta_{n_{ACL}}(t)$ can thus not be observed in Fig.~\ref{fig:roundwsr40} for $p=0.01$. On the other hand, more local cooperators including both ACNLs and SCNLs begin to absorb defectors on the other sides of the local borders (local borders differ from those shortcuts connecting different strategies, mainly consist of local Cs i. e., ACNLs and SCNLs) (see Fig.~\ref{fig:spatialws}(a) and red hollow circles illustrated in middle panels of Fig.~\ref{fig:roundwsr40}) to expand permanently. Correspondingly, we observe obvious positive peaks of $\Delta_{n_{ACNL}}(t)$ and $\Delta_{n_{SCNL}}(t)$ in Fig.~\ref{fig:roundwsr40} (the red hollow circles). Consequently, ACNLs form a stable coexistence with SCNLs (see bottom panels of Fig.~\ref{fig:roundwsr40}); or even give ways to SCNLs because SCNLs are more active than ACNLs in carrying out punishment. After the population reach the sate of full cooperation, it is possible that the majority-like evolution rule would further solidify advantages of SCs; which is supported by the illustration in last subfigure of Fig.~\ref{fig:spatialws}(a).

With increasing $p$, more connections of Cs are allocated to reach the D territories outside C clusters as shortcuts, such that ACLs strike a balance i. e., strong network reciprocity and sanctions against Ds match each other. ACLs' benefits from many other cooperative group members are enough to bear costs of punishment (A large positive peak of $\Delta_{ACL-D}(t)$ can still be found in Fig.~\ref{fig:roundwsr40} for $p=0.09$ which is the optimal randomness of the networks for $\alpha=4.0$). At the same time, they sustain competitive payoffs in comparison to DLs after punishment. By contrast, because of more shortcuts, SCs are also more likely to touch more Ds, and then begin to shift their main attentions from punishment to reserving payoffs. Moreover, ACNLs are not restricted and instead active in expanding, which is affirmed by the sharp positive peak of $\Delta_{n_{ACNL}}(t)$ showed in the first subfigure in middle panels of Fig.~\ref{fig:roundwsr40}. As a result, ACLs absorb the Ds at the other end of shortcuts (these newly transformed Cs actually play a role of 'seed') at a faster speed than SCs (Fig.~\ref{fig:roundwsr40} also present large positive peaks of $\Delta_{ACL-D}(t)$ and $\Delta_{n_{ACL}}(t)$ for $p=0.09$), accompanying a subsequent growth of ACNLs surrounding these 'seeds' (see Fig.~\ref{fig:spatialws}(b)). Then these small clusters centering on the 'seeds' further connect each other to capture the majority of the population. At last, majority-like evolution rule enable ACs to be final winners. 

As $p$ getting larger, owing to appearance of more shortcuts in the network, the number of ACs starts dropping (see bottom panels of Fig.~\ref{fig:roundwsr40}) fast because supports from network reciprocity is lacking, both because ACs exert sanctions too frequently (ACs have more chances to contact more DLs to become ACLs, even worse these Ds can stimulate more interests of ACLs in sanctioning) and even more so because ACLs's payoffs are greatly reduced to be subverted by DLs when punishment effectiveness is not very strong (we could find in top panels of Fig.~\ref{fig:roundwsr40} that ACLs own uncompetitive payoffs for $p=0.13$). Still, dense local connections are more or less kept, which means that SCs rather than ACNLs could possibly sustain high payoffs with the help of network reciprocity (see Fig.~\ref{fig:samer40fp}). Especially, because of the free-riding nature, SCs become the dominator of the whole population again; showing peaks of fractions. As expected, there exists a critical number of shortcuts, above which the dynamics on small-world networks based on 2D RLs is naturally close to what we have observed on RRNs. Weak network reciprocity lead to extinction of Cs, in spite of ACs or SCs (Fig.~\ref{fig:spatialws}(c)); unless the imposed fine $\alpha$ is sufficiently high. This further affirms our interpretations of RRNs playing negative roles in supporting Cs under most parameter conditions. Besides, as a comparison of Fig.~\ref{fig:roundwsr40}, Fig.~\ref{fig:wsround45} in Appendix.~B presents a shorter relaxation time of the system for weak punishment of PGG, attributing to longer term of clustering behaviors of cooperators in this case.   

\begin{figure}[ht!]
\includegraphics[width=0.5\textwidth]{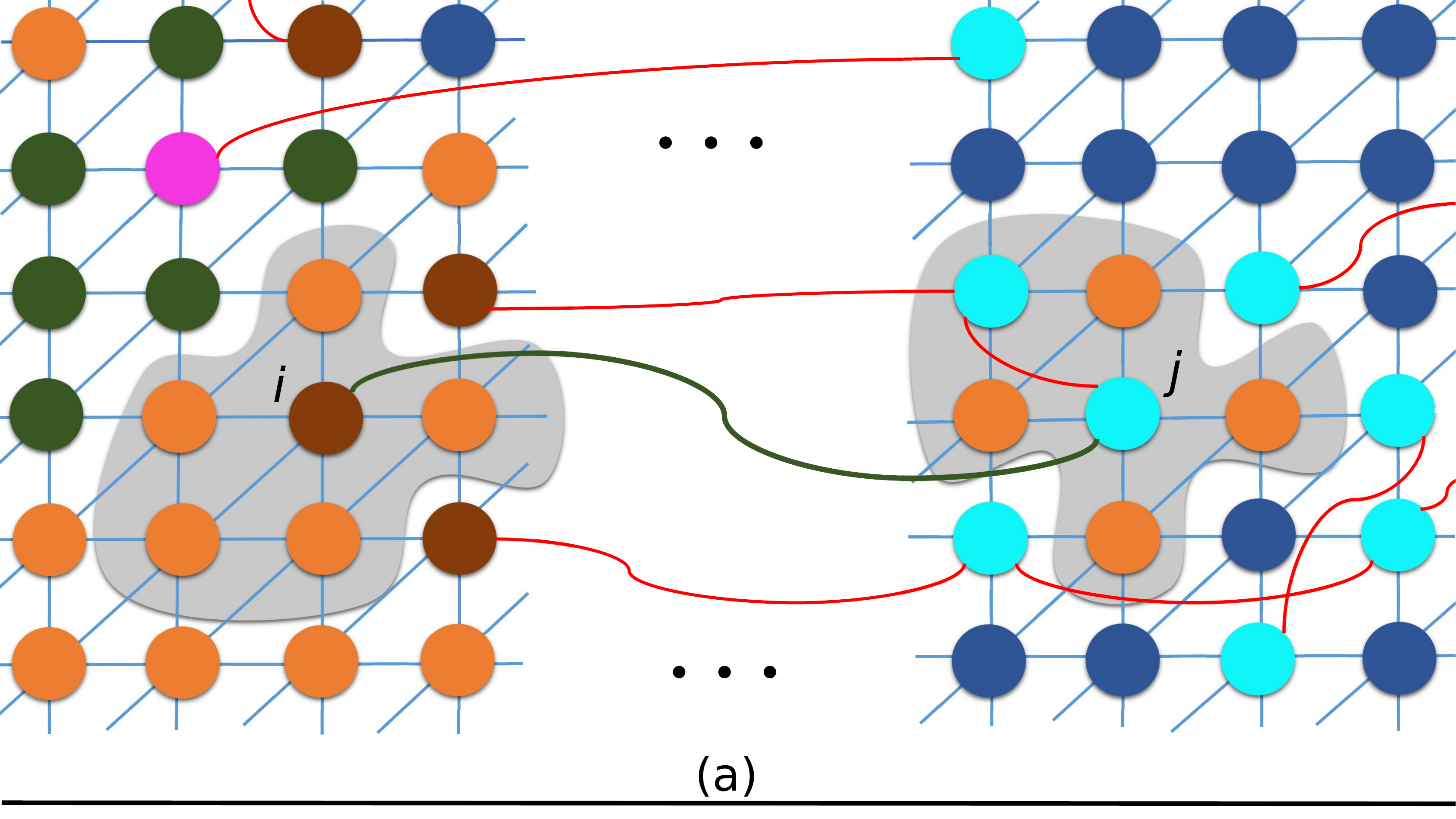}
\vspace*{.5cm}
\includegraphics[width=0.5\textwidth]{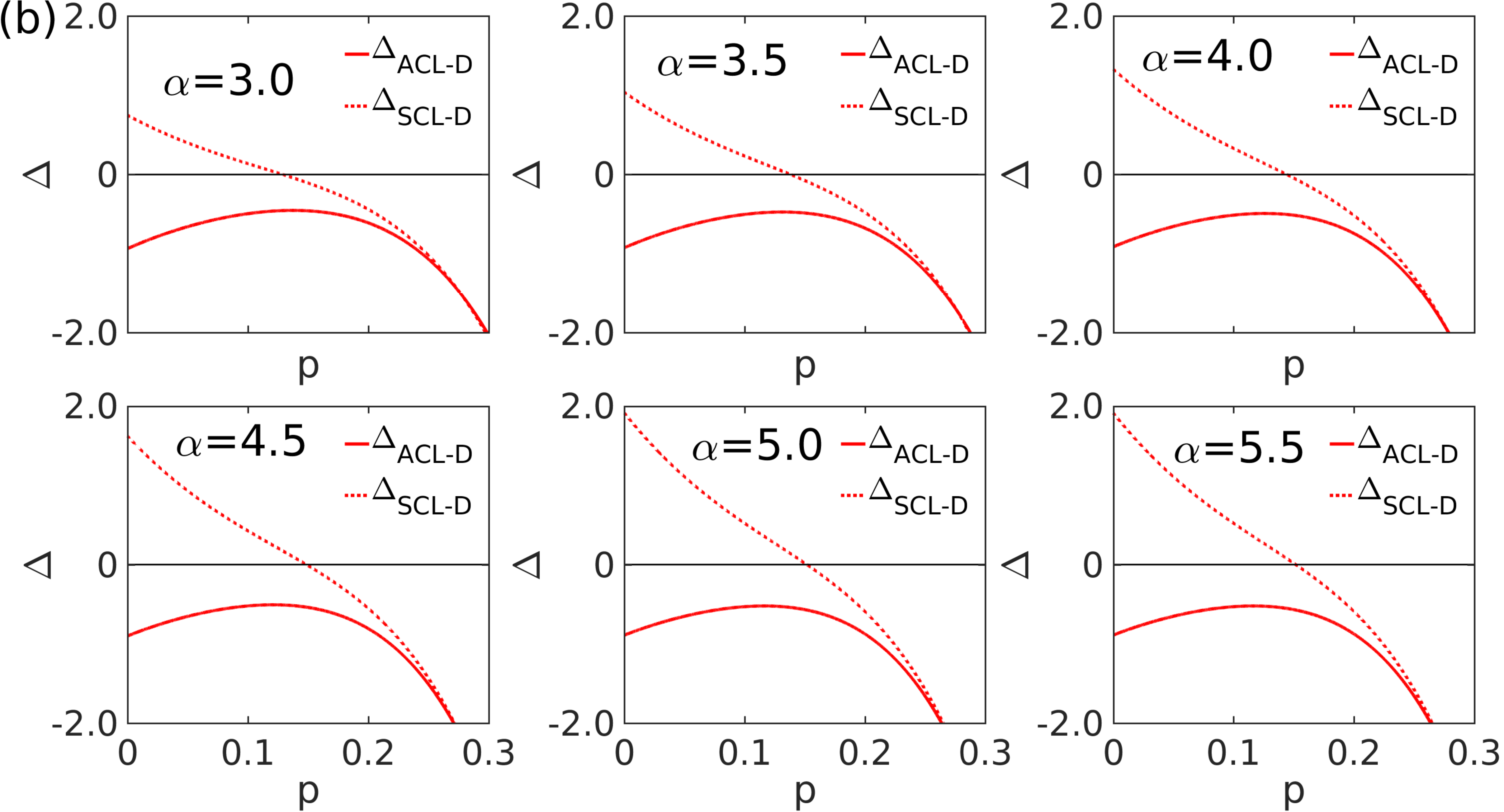}
\caption{(a) Illustration of the extreme situation in which shortcut (the thick green solid curve) originating from the focused altruistic cooperator $i$ is assumed to definitely reach the defector $j$. The red solid curves represent other shortcuts of the networks. Besides the defective neighbors connected by shortcuts, we assume that the local group members centering on $i$ are all ACNLs (SCNLs) if it is an ACL (SCL), and so does the focused defector $j$ at the other end of the shortcut (i.~e. local group members of $j$ are all ACNLs (SCNLs) too). The two grey regions indicate group members of $i$ and $j$, respectively. The individuals of different strategies or types are colored with the same colors as we adopt in Fig.~\ref{fig:spatialws}. (b) The predicted payoff gaps $\Delta_{ACL-D}$ ($\Delta_{SCL-D}$) between ACL (SCL) and its connected defector through the shortcut is plotted as function of randomness $p$ for different values of punishment fines. The employed contact networks are small-world networks based on 2D RLs. The other parameters are taken as $r=4.0$ and $A=0.235$. The analytical solid lines of $\Delta_{ACL-D}$ obviously prove the existence of optimal regions of randomness, regardless of $\alpha$.
}
\label{fig:analysisws40fp}
\end{figure} 
The proposed interpretation for existences of the optimal parameter regions is further supported by our qualitative analysis (i.~e., predicted payoff gaps) based on pair approximation according to the example of extreme situation illustrated in Fig.~\ref{fig:analysisws40fp} (Please see Appendix.~C for details of qualitative analysis). It could be found that predicted payoff gap $\Delta_{ACL-D}$ shows a maximum value near the observed simulated optimal parameter regions. At the same time, it should be noted that $\Delta_{ACL-D}$ are negative, because the most favorable condition for the focused defector is assumed for sake of analysis: all of its local group members are cooperators. However, most defectors are actually in a more disadvantaged evolutionary position, since most of their local group members are instead defective guys, which is also supported by the snapshots illustrated in Fig.~\ref{fig:spatialws}. As a result, ACLs can own higher payoffs than their defective neighbors after sanctions, especially in the optimal parameter regions. After all, the qualitative analysis successfully identify that ACLs play a key role in facilitating prevalence of ACs in the population, by suggesting the most likehood of ACLs' expansion which is positively related to $\Delta_{ACL-D}$. That is in accordance our interpretation.

\begin{figure}[ht!]
\includegraphics[width=0.5\textwidth]{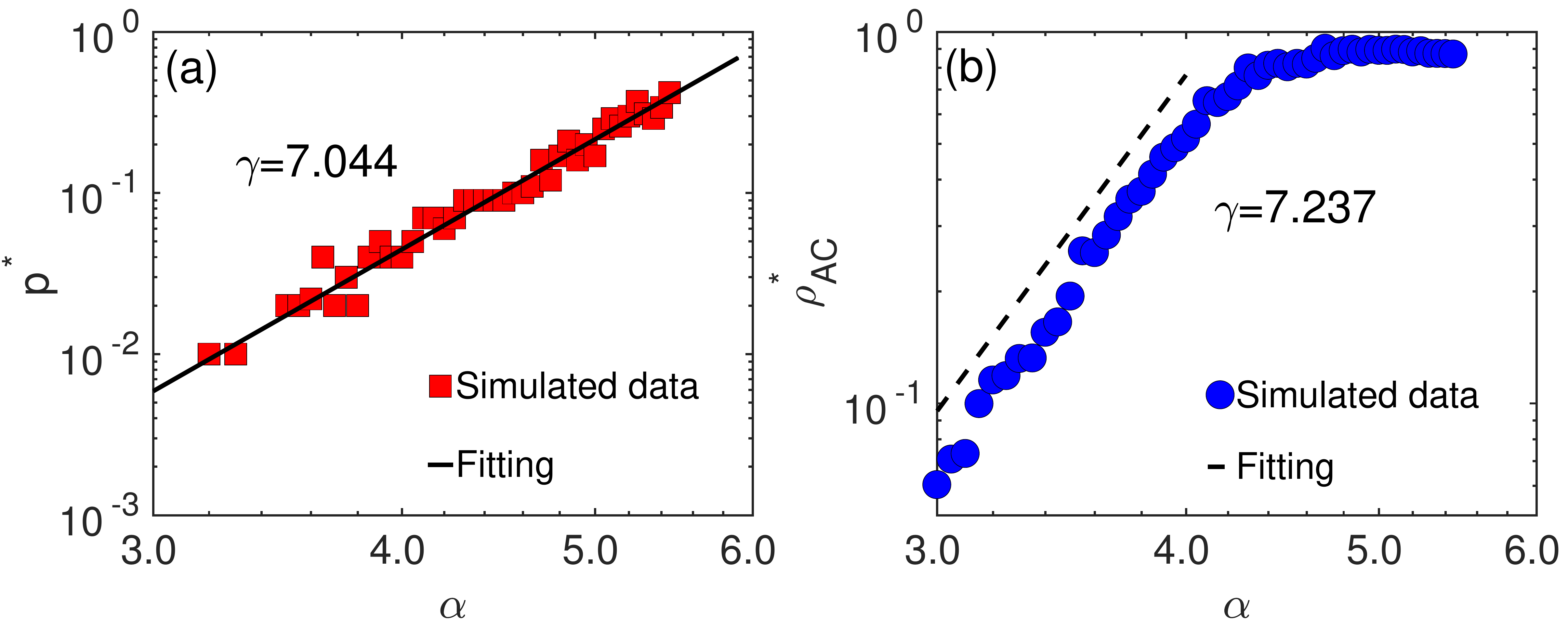}
\caption{The dependence of (a) optimal randomness $p^{*}$ and (b) maximum population fraction of ACs $\rho^{*}_{AC}$ on punishment fine; which is fitted to a power-law increase of the forms $ p^{*} \sim \alpha^{\gamma}$ and $\rho^{*}_{AC}\sim \alpha^{\gamma}$, respectively. The employed networks are the small-world networks based on 2D RLs. The other parameters are taken as $r=4.0$ and $A=0.235$. In (a), the dark solid line indicates the fitted relationship between $p^{*}$ and $\alpha$; while in (b) the dark dashed line indicates the power fit of the relationship between $\alpha$ and $f^{*}_{AC}$ in region $3.0<\alpha<4.0$. In detail, $\rho^{*}_{AC}$ is maximum population fraction of ACs that the system could reach for each $\alpha$ (i~e. $\rho_{AC}$ at $p^{*}$). In both cases, the fitted relationships are statistically significant ((a) $R^{2} = 0.9671$ and $P < 0.01$; (b) $R^{2} = 0.9926$ and $P<0.01$).  
}
\label{fig:alphap40}
\end{figure}
The dependence of optimal randomness and maximum population fractions of AC on punishment fines, respectively depicted in the subfigure of Fig.~\ref{fig:alphap40}, allows us to comprehensively understand how the small-world topology governs the performance of pool punishment in changing optimal randomness and facilitating ACs. We see clearly in Fig.~\ref{fig:alphap40}(a) that optimal randomness of the networks increase with punishment fine in a power-law-like way. It is also true for behavior of maximum population fraction of AC; until $\alpha\simeq 4.0$ above which strong punishment can do nothing in further promoting ACs because the system begins to reach saturation in ACs. This interesting and important phenomenon tells us that stronger pool punishment is not always equivalent to higher levels of altruistic cooperation in presence of two modes of conformity feedback, which is more or less against people's common sense.

To sum up, in the parameter regions of strong network reciprocity, punishment has a positive impact on the evolution of cooperation. The desired outcome reveals great potentials of the small-world networks based on 2D RLs in overcoming the first-order and the second-order dilemmas. As we point out, both Fig.~\ref{fig:samer40fp}, Fig.~\ref{fig:roundwsr40}, Fig.~\ref{fig:spatialws} and Fig.~\ref{fig:analysisws40fp} illustrate clearly that shortcuts introduced by randomness $p$ play a decisive role in reaching the correct evolutionary outcome from a random initial state.

\section{Conclusion and Discussion}
\label{sec:discuss}
In this paper we have proposed an evolutionary game model considering two converse feedback modes of conformity that exerting strong influence on punishment of cooperators, in framework of PGG model. Next, we have proceeded detailed investigations on the implications of the model on three kinds of networks: regular lattice networks (RLs), regular random networks (RRNs) and small-world networks based on 2D RLs; aiming to find an optimal topology to overcome the first-order and the second-order social dilemmas. In cases of respective networks, we have shown how network spatial structures, feedback sensitivity of individuals, the effectiveness of punishment and synergistic effects of cooperation govern the evolution of cooperators (Cs), altruistic cooperators (ACs) and selfish cooperators (SCs). Agent-based Monte Carlo (MC) method employed in the simulations gives rich and interesting outcomes, which are also supported and confirmed by the theoretical analysis on basis of mean-field theory. 

In more detail, in the second part of the paper we firstly explore the evolutionary dynamics on RLs. Indeed, we find that RLs overall promote cooperation especially selfish cooperation due to strong network reciprocity from abundant local connections; which is also in accordance with the conclusions from previous studies concerning evolutionary dynamics on networks~\cite{szabo2007,nowak2006,szolnoki2009}. Nevertheless, altruistic cooperation is in a particularly vulnerable situation for the general case except that punishment fine is rather large and $A$ is intermediate. This implies that the second-order dilemma persists as a trouble. In the analytical way employing a set of equations of the replicator dynamics, we could derive a mathematical expression of the critical boundary between C phase and D phase, which turns out to depend on degree of each node in the RLs, feedback sensitivity of individuals and synergistic effects of cooperation. We find a nice agreement between simulation results and analytical predictions. 

In subsequent case of RRNs, things become different and perhaps more negative. Cs are completely suppressed by Ds in most $A-\alpha$ parameter region because of weak network reciprocity from less overlaps among PGG groups, however, strong effectiveness of punishment and mediate conformity feedback sensitivity may together contribute to the dominance of ACs. In other words, ACs have an advantage over other guys in the limited cases while whole cooperator population overall suffer from free riding of defectors. The results for RRNs suggest that the evolutionary dynamics is more dependent on the control parameters such as punishment fine, synergy factor and conformity feedback sensitivity of individuals; because the system is closed to well-mixed situation. Besides, fractions of ACs show sharper fluctuations at the critical boundaries, we thus instead use one estimated constant fraction of ACs in analysis. Coincidence between simulated phase areas and predicted boundaries proves that our semi-analytical method is useful. Taken together, although RLs and RRNs themselves are not best choices for ACs , the results on the two networks correctly lead us to precise characteristics of the optimal topology we are looking for: owning both dense local connections and mediate shortcuts.   

Especially and naturally, a very interesting point in this respect regards what happens on the small-world networks; since this topology owns both dense local connections and shortcuts originating from random reconnections. In such a case, quite remarkably, simulation results show that there indeed exists a moderate optimal parameter region in terms of randomness of the networks, in which altruistic cooperators (ACs) capture the majority of the population. This is in agreement with the qualitative analysis assuming an extreme situation that the shortcuts from an altruistic cooperator (selfish cooperator) would point to a defector without doubt. For more details, microscopic mechanism behind the reported evolutionary outcomes can be explained by the comparison between strong network reciprocity and considerable but favorable chances to contact Ds through shortcuts. A balance between the two factors can help ACs outperform defectors, and further prevail over SCs. The results prove that the small-world networks based on 2D RLs can not only help cooperators successfully suppress defectors by means of strong network reciprocity stemming from dense local connections, but also provide sufficient contacts between ACs and Ds to facilitate the expansion and prevalence of ACs in presence of strong effectiveness of punishment. In a narrow region of the large $p$, SCs can instead be in a dominated position. But apart from this, as network reciprocity weakening with more shortcuts, punishment fails and cooperation is evolutionary unsuccessful. These results are also robust to changes of synergistic effects of cooperation. Furthermore, we find an interesting phenomenon, that is on the small-world networks the maximum population of ACs is actually bounded, i.~e., ACs can not capture overall population; no matter how strong pool punishment is. In summary, small-world networks based on 2D RLs turn out to be one optimal choice to sustain public goods by alleviating both the first-order and the second-order dilemmas.  

In comparison to previous researches~\cite{fu2007,santos2008,chen2014}, our study proposes a new and more realistic theoretical framework to curve the evolutionary dynamics capturing interplay between conformity preference and punishment behavior. This model as well as the analytical method can be employed  by future further investigations concerning evolutionary game dynamics, as a basic theoretical tool. Our study also reveals conditions that favor cooperation especially altruistic cooperation: sufficiently high multiplication factor of the public goods game, mediate conformity feedback sensitivity of individuals, strong effectiveness of punishment and the most important factor: small-world-like social connections. This provides an initial guide for governments or enterprises on how to sustain or even facilitate social public goods, or on how to improve return rate.

It is also worth noticing that in this study we do not consider the situation that strategy states of group members could exert influence on changes of individual's strategy~\cite{cui2013} or antisocial punishment~\cite{szolnoki2017}. In the light of this fact, it is important and interesting to extend our current theory of cooperation in spatial PGG with antisocial punishment or the option that shift of individuals'  strategies are also influenced by conformity.

\section*{Acknowledgments}
This work was supported by China Postdoctoral Science Foundation No.~2015M582532 and by the National Natural Science Foundation of China (Grant Nos.~61433014, 61473060, 11575072 and 11475074).

\section*{Appendix A}
\label{sec:appendixa}
In both genetic evolutions, the evolutionary process can be analytically described by a set of equations called the replicator dynamics~\cite{weibull1997}. The evolutionary dynamics of the studied system could thus be determined by the following replicator equation: 
\begin{eqnarray}
\frac{df}{dt}=f(1-f)(\Pi_{X}-\Pi_{D})
\label{eq:replicator}
\end{eqnarray}
where $f$ is the fraction of all the cooperators in the population. $\Pi_{X}=x\Pi_{P}+(1-x)\Pi_{C}$ represents the average payoff of all the cooperators. $x$ is the expected probability that a cooperator become a punisher in the group. $\Pi_{C}$, $\Pi_{P}$ and $\Pi_{D}$ represent the average payoffs of cooperators, punishers and defectors, respectively.

To theoretically investigate the evolution of cooperation, it is assumed that in each round of the game an interaction group is assembled by randomly selecting $n$ individuals from the population. Accordingly, we can obtain the expressions of $\Pi_{C}$, $\Pi_{P}$ and $\Pi_{D}$ respectively:
\begin{widetext}
\begin{eqnarray}
\label{eq:threepayoff1}
\Pi_{P}  & = & \sum\limits_{i=0}^{n-1}\dbinom{n-1}{i}f^{i}(1-f)^{n-1-i} \sum\limits_{j=0}^{i}\dbinom{i}{j}g^{j}(1-g)^{i-j}[\frac{r(i+1)}{n}-1-\frac{\alpha(n-1-i)}{j+1}],\\
\label{eq:threepayoff2}
\Pi_{C} & = & \sum\limits_{i=0}^{n-1}\dbinom{n-1}{i}f^{i}(1-f)^{n-1-i} \sum\limits_{j=0}^{i}\dbinom{i}{j}g^{j}(1-g)^{i-j}[\frac{r(i+1)}{n}-1],\\
\label{eq:threepayoff3}
\Pi_{D} & = & \sum\limits_{i=0}^{n-1}\dbinom{n-1}{i}f^{i}(1-f)^{n-1-i}\sum\limits_{j=1}^{i}\dbinom{i}{j}g^{j}(1-g)^{i-j}[\frac{ri}{n}-\alpha]+\sum\limits_{i=0}^{n-1}\dbinom{n-1}{i}f^{i}(1-f)^{n-1-i}(1-g)^{i}\frac{ir}{n},
\end{eqnarray}
\end{widetext} 
where $g(p_{s}(i),p_{a}(i))$ is the probability that a randomly selected cooperator becomes a punisher, as the function of number of cooperators $i$ in the group. For simplicity, it is assumed that the relationship $g=(1-y)p_{s}+yp_{a}$ is hold; where $y$ ($1-y$) represents the weight of contribution of ACs (SCs) to the probability $g$. Based on Eq.~\ref{eq:asfunztion1}, the following relationship is obtained:
\begin{eqnarray}
g(i) & = & y+(1-2y)p_{s}(i) \notag \\
& = & y+(1-2y)\frac{Ai}{n}
\label{eq:pfunction}
\end{eqnarray}
Furthermore, combing Eq.~\ref{eq:threepayoff1}, Eq.~\ref{eq:threepayoff2} and Eq.~\ref{eq:threepayoff3}, we obtain:
\begin{eqnarray}
h & = & \Pi_{X}-\Pi_{D} \notag \\ 
& = & \frac{r}{n}-1+\alpha(1-\phi-\psi)
\label{eq:gfunction}
\end{eqnarray}
where $h$ is the function of parameters $A$, $r$, $x$ and $y$. $\phi=\sum\limits_{i=0}^{n-1}\dbinom{n-1}{i}f^{i}(1-f)^{n-1-i}(1-g)^{i}$,$\psi= x\sum\limits_{i=0}^{n-1}\dbinom{n-1}{i}f^{i}(1-f)^{n-1-i}  \sum\limits_{j=0}^{i}\dbinom{i}{j}g^{j}(1-g)^{i-j}[\frac{n-1-i}{j+1}]$.

Eq.~\ref{eq:replicator} could be translated into:
\begin{eqnarray}
\frac{df}{dt}=f(1-f)h(A,r,x,y)
\label{eq:replicator2}
\end{eqnarray}
It could be found that there are two stable boundary equilibria $f=0$ and $f=1$, and one unstable interior equilibria which is determined by $h(A,r,x,y)=0$. Actually, $g=g(A,\alpha)$, which means that $g$ is fully determined by $A$ and $\alpha$, even though the functional form is not clear. Therefore, $g$ could be considered as a constant, in addition to $x=g$. Therefore Eq.~\ref{eq:replicator2} could be reduced to:
\begin{eqnarray}
h=(-1+\frac{r}{n})+a[1-(1-gf)^{n-1}](1-\frac{1-f}{f})
\label{eq:gsimple}
\end{eqnarray}
for $\frac{df}{dt}=0$.

Next, we focused on theoretical predictions of the boundaries. Fig.~\ref{fig:hexAalpha} shows that the selfish cooperators are dominated near the boundary, so we let $f=f_{2}$ ($f_{2}$ represents the fractions of selfish cooperators) and $g(i)=p_{s}(i)=A\frac{i}{n}$. Based on Eq.~\ref{eq:threepayoff1}, Eq.~\ref{eq:threepayoff2}, Eq.~\ref{eq:threepayoff3} and Eq.~\ref{eq:gfunction}, we have:
\begin{eqnarray}
\alpha(A,g,f_{2})=\frac{1-r/n}{1-\psi(A,g,f_{2})-\phi(A,f_{2})}
\label{eq:thresholdfunction}
\end{eqnarray}
where $g=A\frac{nf_{2}}{n}=Af_{2}$ in the point of full mean-field theory.
The stability analysis of the system implies that $f_{2}=1$ on one side of the boundary, so theoretical predictions could be obtained for RLs by directly setting $f_{2}=1$. However, because of small synergy factor, it could be found that $f_{2}<1$ near the boundary due to existence of interior equilibria shown in Fig~\ref{fig:hexAalpha}(a). As a result, semi-theoretical analysis by adjusting the value of $f_{2}$ instead provide more accurate prediction.

In case of RRNs, Fig.~\ref{fig:rraalpha} shows that ACs are dominated near the boundary between full-D phase and full-C phase, so we instead let $f=f_{1}$ ($f_{1}$ represents the fractions of ACs near the boundary) and $p(i)=p_{a}(i)=A-A\frac{i}{n}$. Based on Eq.~\ref{eq:thresholdfunction}, we have:
\begin{eqnarray}
\alpha(A,g,f_{1})=\frac{1-r/n}{1-\psi(A,g,f_{1})-\phi(A,f_{1})}
\label{eq:thresholdfunction}
\end{eqnarray}
where $g=p_{a}(i)=A(1-\frac{i}{n})$ in the point of full mean-field theory. Since ACs and SCs coexist together near the boundary (i.e., $f_{1}\not\approx 1.0$), only semi-theoretical analysis could be get by estimating the values of $f_{1}$ in Eq.~\ref{eq:thresholdfunction}.

\section*{Appendix B}
\label{sec:appendixb}

\begin{figure}[ht!]
\includegraphics[width=0.5\textwidth]{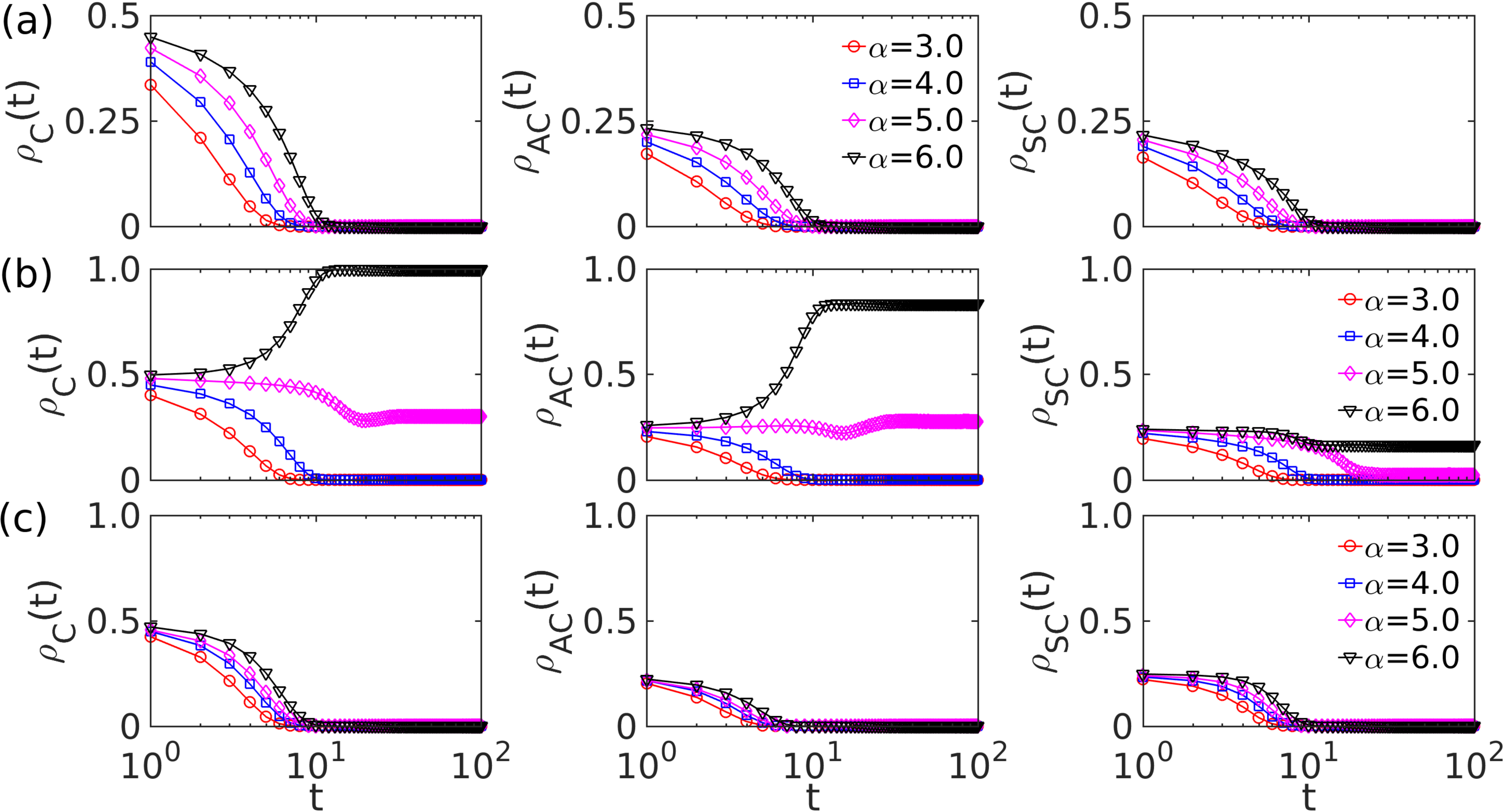}
\caption{Evolution of fractions of three different types of cooperators for different punishment fines on RRs. The values of feedback sensitivity are $A=0.03$ (a) $A=0.20$ (b) and $A=0.55$ (c); respectively. The other parameter is $r=4.0$.}
\label{fig:rrnround}
\end{figure}

\begin{figure}[ht!]
\includegraphics[width=0.5\textwidth]{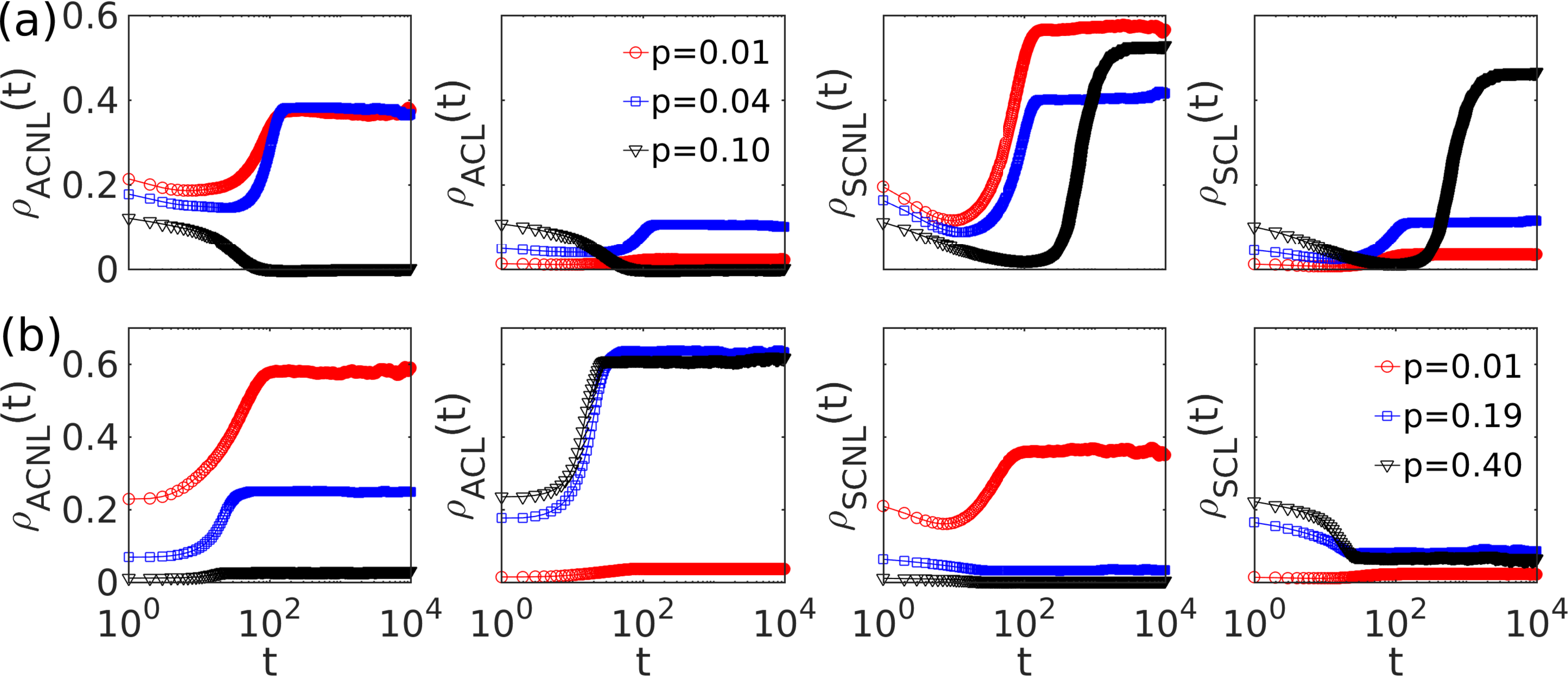}
\caption{Evolution of fractions of four different types of cooperators for different randomness on the small-world networks base on 2D RLs. The values of punishment fine are $\alpha=4.0$ (a) and $\alpha=5.0$ (b); respectively. The other parameters are $r=4.0$ and $A=0.235$.}
\label{fig:wsround45}
\end{figure}

\begin{figure}[ht!]
\includegraphics[width=0.5\textwidth]{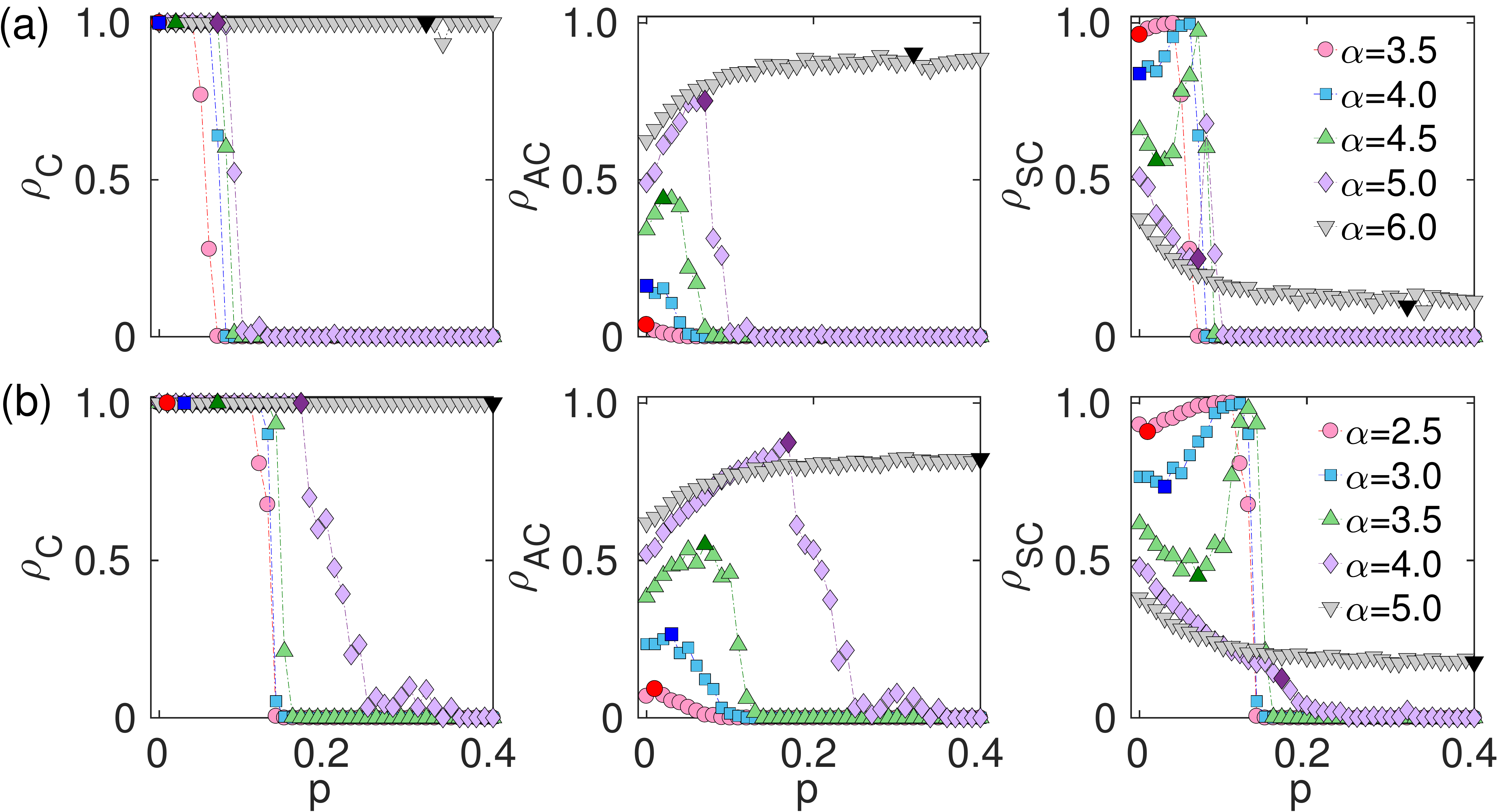}
\caption{Fractions of three different strategies: $\rho_{C}$,  $\rho_{AC}$ and $\rho_{SC}$, plotted versus $p$ for two different values synergy factor: $r=3.5$ (a) and $r=4.5$ (b). The value of feedback sensitivity is $A=0.235$. The employed networks are the small-world networks based on 2D RLs. The dark marker for each $\alpha$ corresponds to the position of optimal randomness where the population of ACs is maximum.
}
\label{fig:2dws3545}
\end{figure}

\section*{Appendix C}
\label{sec:appendixc}
According to the illustration in Fig.~\ref{fig:analysisws40fp}(a), the payoffs of the focused altruistic cooperator and its connected defector are
\begin{widetext}
\begin{eqnarray}
\Pi_{i}  & = & \sum\limits_{l=0}^{n-2}\dbinom{n-2}{l}\Large(\frac{p}{2}\Large)^{l}(1-\Large(\frac{p}{2}\Large)^{n-2-l})\Large[ p_{a}(\frac{r(n-l-1)}{n}-1.0-\frac{\alpha(l+1)}{(n-l-1)p_{a}})+(1-p_{a})(\frac{r(n-l-1)}{n}-1.0) \Large],\\
\Pi_{j}  & = & \sum\limits_{l=0}^{n-2}\dbinom{n-2}{l}\Large(\frac{p}{2}\Large)^{l}(1-\Large(\frac{p}{2}\Large)^{n-2-l})\Large[(1-(1-p_{a})^{n-l-1})(\frac{r(n-l-1)}{n}-\alpha)+ (1-p_{a})^{n-l-1}(\frac{r(n-l-1)}{n})\Large],
\label{eq:payacldl}
\end{eqnarray}
\end{widetext} where $p_{a}=A\frac{i+1}{n}$ and $l$ represent the number of shortcuts except the one between $i$ and $j$. Based on our assumption, the number of defective group members is $l$ for $j$, and $l+1$ for $i$. Eq.~\ref{eq:payacldl} naturally gives the payoff gap between $i$ and $j$ 
\begin{eqnarray}
\Delta_{ACL-D}=\Pi_{i}-\Pi_{j}
\label{eq:paygapacl}
\end{eqnarray}, which is function of $p$. Similarly, we could obtain the payoff gap between SCL and its connected defector $\Delta_{SCL-D}$ if $p_{a}$ in Eq.~\ref{eq:payacldl} is replaced with $p_{s}=A\frac{n-l-1}{n}$.

\bibliography{multireference}

\begin{thebibliography}{37}%
\makeatletter
\providecommand \@ifxundefined [1]{%
 \@ifx{#1\undefined}
}%
\providecommand \@ifnum [1]{%
 \ifnum #1\expandafter \@firstoftwo
 \else \expandafter \@secondoftwo
 \fi
}%
\providecommand \@ifx [1]{%
 \ifx #1\expandafter \@firstoftwo
 \else \expandafter \@secondoftwo
 \fi
}%
\providecommand \natexlab [1]{#1}%
\providecommand \enquote  [1]{``#1''}%
\providecommand \bibnamefont  [1]{#1}%
\providecommand \bibfnamefont [1]{#1}%
\providecommand \citenamefont [1]{#1}%
\providecommand \href@noop [0]{\@secondoftwo}%
\providecommand \href [0]{\begingroup \@sanitize@url \@href}%
\providecommand \@href[1]{\@@startlink{#1}\@@href}%
\providecommand \@@href[1]{\endgroup#1\@@endlink}%
\providecommand \@sanitize@url [0]{\catcode `\\12\catcode `\$12\catcode
  `\&12\catcode `\#12\catcode `\^12\catcode `\_12\catcode `\%12\relax}%
\providecommand \@@startlink[1]{}%
\providecommand \@@endlink[0]{}%
\providecommand \url  [0]{\begingroup\@sanitize@url \@url }%
\providecommand \@url [1]{\endgroup\@href {#1}{\urlprefix }}%
\providecommand \urlprefix  [0]{URL }%
\providecommand \Eprint [0]{\href }%
\providecommand \doibase [0]{http://dx.doi.org/}%
\providecommand \selectlanguage [0]{\@gobble}%
\providecommand \bibinfo  [0]{\@secondoftwo}%
\providecommand \bibfield  [0]{\@secondoftwo}%
\providecommand \translation [1]{[#1]}%
\providecommand \BibitemOpen [0]{}%
\providecommand \bibitemStop [0]{}%
\providecommand \bibitemNoStop [0]{.\EOS\space}%
\providecommand \EOS [0]{\spacefactor3000\relax}%
\providecommand \BibitemShut  [1]{\csname bibitem#1\endcsname}%
\let\auto@bib@innerbib\@empty
\bibitem [{\citenamefont {Foster}\ \emph {et~al.}(2006)\citenamefont {Foster},
  \citenamefont {Wenseleers}, \citenamefont {Ratnieks},\ and\ \citenamefont
  {Queller}}]{foster2006}%
  \BibitemOpen
  \bibfield  {author} {\bibinfo {author} {\bibfnamefont {K.~R.}\ \bibnamefont
  {Foster}}, \bibinfo {author} {\bibfnamefont {T.}~\bibnamefont {Wenseleers}},
  \bibinfo {author} {\bibfnamefont {F.~L.}\ \bibnamefont {Ratnieks}}, \ and\
  \bibinfo {author} {\bibfnamefont {D.~C.}\ \bibnamefont {Queller}},\ }\href
  {\doibase 110.1016/j.tree.2006.08.010} {\bibfield  {journal} {\bibinfo
  {journal} {Trends Ecol. Evol.}\ }\textbf {\bibinfo {volume} {21}},\ \bibinfo
  {pages} {599} (\bibinfo {year} {2006})}\BibitemShut {NoStop}%
\bibitem [{\citenamefont {Alexander}(1987)}]{alexander1987}%
  \BibitemOpen
  \bibfield  {author} {\bibinfo {author} {\bibfnamefont {R.~D.}\ \bibnamefont
  {Alexander}},\ }\href@noop {} {\emph {\bibinfo {title} {The biology of moral
  systems}}}\ (\bibinfo  {publisher} {Transaction Publishers},\ \bibinfo
  {address} {New Jersey, USA},\ \bibinfo {year} {1987})\BibitemShut {NoStop}%
\bibitem [{\citenamefont {Axelrod}\ and\ \citenamefont
  {Axelrod}(1984)}]{axelrod1984}%
  \BibitemOpen
  \bibfield  {author} {\bibinfo {author} {\bibfnamefont {R.}~\bibnamefont
  {Axelrod}}\ and\ \bibinfo {author} {\bibfnamefont {R.~M.}\ \bibnamefont
  {Axelrod}},\ }\href@noop {} {\emph {\bibinfo {title} {The evolution of
  cooperation}}},\ Vol.\ \bibinfo {volume} {5145}\ (\bibinfo  {publisher}
  {Basic Books (AZ)},\ \bibinfo {year} {1984})\BibitemShut {NoStop}%
\bibitem [{\citenamefont {Nowak}(2006)}]{nowak2006}%
  \BibitemOpen
  \bibfield  {author} {\bibinfo {author} {\bibfnamefont {M.~A.}\ \bibnamefont
  {Nowak}},\ }\href {\doibase 10.1126/science.1133755} {\bibfield  {journal}
  {\bibinfo  {journal} {Science}\ }\textbf {\bibinfo {volume} {314}},\ \bibinfo
  {pages} {1560} (\bibinfo {year} {2006})}\BibitemShut {NoStop}%
\bibitem [{\citenamefont {Fehr}\ and\ \citenamefont
  {G{\"a}chter}(2000)}]{fehr2000}%
  \BibitemOpen
  \bibfield  {author} {\bibinfo {author} {\bibfnamefont {E.}~\bibnamefont
  {Fehr}}\ and\ \bibinfo {author} {\bibfnamefont {S.}~\bibnamefont
  {G{\"a}chter}},\ }\href {\doibase 10.1257/aer.90.4.980} {\bibfield  {journal}
  {\bibinfo  {journal} {Am. Econ. Rev.}\ }\textbf {\bibinfo {volume} {90}},\
  \bibinfo {pages} {980} (\bibinfo {year} {2000})}\BibitemShut {NoStop}%
\bibitem [{\citenamefont {Fehr}\ and\ \citenamefont
  {G{\"a}chter}(2002)}]{fehr2002}%
  \BibitemOpen
  \bibfield  {author} {\bibinfo {author} {\bibfnamefont {E.}~\bibnamefont
  {Fehr}}\ and\ \bibinfo {author} {\bibfnamefont {S.}~\bibnamefont
  {G{\"a}chter}},\ }\href {\doibase 10.1038/415137a} {\bibfield  {journal}
  {\bibinfo  {journal} {Nature}\ }\textbf {\bibinfo {volume} {415}},\ \bibinfo
  {pages} {137} (\bibinfo {year} {2002})}\BibitemShut {NoStop}%
\bibitem [{\citenamefont {G{\"a}chter}\ \emph {et~al.}(2008)\citenamefont
  {G{\"a}chter}, \citenamefont {Renner},\ and\ \citenamefont
  {Sefton}}]{gachter2008}%
  \BibitemOpen
  \bibfield  {author} {\bibinfo {author} {\bibfnamefont {S.}~\bibnamefont
  {G{\"a}chter}}, \bibinfo {author} {\bibfnamefont {E.}~\bibnamefont {Renner}},
  \ and\ \bibinfo {author} {\bibfnamefont {M.}~\bibnamefont {Sefton}},\ }\href
  {\doibase 10.1126/science.1164744} {\bibfield  {journal} {\bibinfo  {journal}
  {Science}\ }\textbf {\bibinfo {volume} {322}},\ \bibinfo {pages} {1510}
  (\bibinfo {year} {2008})}\BibitemShut {NoStop}%
\bibitem [{\citenamefont {Herrmann}\ \emph {et~al.}(2008)\citenamefont
  {Herrmann}, \citenamefont {Th{\"o}ni},\ and\ \citenamefont
  {G{\"a}chter}}]{herrmann2008}%
  \BibitemOpen
  \bibfield  {author} {\bibinfo {author} {\bibfnamefont {B.}~\bibnamefont
  {Herrmann}}, \bibinfo {author} {\bibfnamefont {C.}~\bibnamefont {Th{\"o}ni}},
  \ and\ \bibinfo {author} {\bibfnamefont {S.}~\bibnamefont {G{\"a}chter}},\
  }\href {\doibase 10.1126/science.1153808} {\bibfield  {journal} {\bibinfo
  {journal} {Science}\ }\textbf {\bibinfo {volume} {319}},\ \bibinfo {pages}
  {1362} (\bibinfo {year} {2008})}\BibitemShut {NoStop}%
\bibitem [{\citenamefont {Panchanathan}\ and\ \citenamefont
  {Boyd}(2004)}]{panchanathan2004}%
  \BibitemOpen
  \bibfield  {author} {\bibinfo {author} {\bibfnamefont {K.}~\bibnamefont
  {Panchanathan}}\ and\ \bibinfo {author} {\bibfnamefont {R.}~\bibnamefont
  {Boyd}},\ }\href {\doibase 10.1038/nature02978} {\bibfield  {journal}
  {\bibinfo  {journal} {Nature}\ }\textbf {\bibinfo {volume} {432}},\ \bibinfo
  {pages} {499} (\bibinfo {year} {2004})}\BibitemShut {NoStop}%
\bibitem [{\citenamefont {Sigmund}\ \emph {et~al.}(2010)\citenamefont
  {Sigmund}, \citenamefont {De}, \citenamefont {Traulsen},\ and\ \citenamefont
  {Hauert}}]{sigmund2010}%
  \BibitemOpen
  \bibfield  {author} {\bibinfo {author} {\bibfnamefont {K.}~\bibnamefont
  {Sigmund}}, \bibinfo {author} {\bibfnamefont {S.~H.}\ \bibnamefont {De}},
  \bibinfo {author} {\bibfnamefont {A.}~\bibnamefont {Traulsen}}, \ and\
  \bibinfo {author} {\bibfnamefont {C.}~\bibnamefont {Hauert}},\ }\href
  {\doibase 10.1038/nature09203} {\bibfield  {journal} {\bibinfo  {journal}
  {Nature}\ }\textbf {\bibinfo {volume} {466}},\ \bibinfo {pages} {861}
  (\bibinfo {year} {2010})}\BibitemShut {NoStop}%
\bibitem [{\citenamefont {Perc}(2012)}]{perc2012}%
  \BibitemOpen
  \bibfield  {author} {\bibinfo {author} {\bibfnamefont {M.}~\bibnamefont
  {Perc}},\ }\href {http://dx.doi.org/10.1038/srep00344} {\bibfield  {journal}
  {\bibinfo  {journal} {Sci. Rep.}\ }\textbf {\bibinfo {volume} {2}},\ \bibinfo
  {pages} {344} (\bibinfo {year} {2012})}\BibitemShut {NoStop}%
\bibitem [{\citenamefont {Hauert}\ \emph {et~al.}(2002)\citenamefont {Hauert},
  \citenamefont {De~Monte}, \citenamefont {Hofbauer},\ and\ \citenamefont
  {Sigmund}}]{hauert2002}%
  \BibitemOpen
  \bibfield  {author} {\bibinfo {author} {\bibfnamefont {C.}~\bibnamefont
  {Hauert}}, \bibinfo {author} {\bibfnamefont {S.}~\bibnamefont {De~Monte}},
  \bibinfo {author} {\bibfnamefont {J.}~\bibnamefont {Hofbauer}}, \ and\
  \bibinfo {author} {\bibfnamefont {K.}~\bibnamefont {Sigmund}},\ }\href
  {\doibase 10.1126/science.1070582} {\bibfield  {journal} {\bibinfo  {journal}
  {Science}\ }\textbf {\bibinfo {volume} {296}},\ \bibinfo {pages} {1129}
  (\bibinfo {year} {2002})}\BibitemShut {NoStop}%
\bibitem [{\citenamefont {Mathew}\ and\ \citenamefont
  {Boyd}(2009)}]{mathew2009}%
  \BibitemOpen
  \bibfield  {author} {\bibinfo {author} {\bibfnamefont {S.}~\bibnamefont
  {Mathew}}\ and\ \bibinfo {author} {\bibfnamefont {R.}~\bibnamefont {Boyd}},\
  }\href {\doibase 10.1098/rspb.2008.1623} {\bibfield  {journal} {\bibinfo
  {journal} {Proc. Royal Soc. B}\ }\textbf {\bibinfo {volume} {276}},\ \bibinfo
  {pages} {1167} (\bibinfo {year} {2009})}\BibitemShut {NoStop}%
\bibitem [{\citenamefont {Boyd}\ \emph {et~al.}(2010)\citenamefont {Boyd},
  \citenamefont {Gintis},\ and\ \citenamefont {Bowles}}]{boyd2010}%
  \BibitemOpen
  \bibfield  {author} {\bibinfo {author} {\bibfnamefont {R.}~\bibnamefont
  {Boyd}}, \bibinfo {author} {\bibfnamefont {H.}~\bibnamefont {Gintis}}, \ and\
  \bibinfo {author} {\bibfnamefont {S.}~\bibnamefont {Bowles}},\ }\href
  {\doibase 10.1126/science.1183665} {\bibfield  {journal} {\bibinfo  {journal}
  {Science}\ }\textbf {\bibinfo {volume} {328}},\ \bibinfo {pages} {617}
  (\bibinfo {year} {2010})}\BibitemShut {NoStop}%
\bibitem [{\citenamefont {Cui}\ and\ \citenamefont {Wu}(2014)}]{cui2014}%
  \BibitemOpen
  \bibfield  {author} {\bibinfo {author} {\bibfnamefont {P.}~\bibnamefont
  {Cui}}\ and\ \bibinfo {author} {\bibfnamefont {Z.-X.}\ \bibnamefont {Wu}},\
  }\href {\doibase 10.1016/j.jtbi.2014.07.021} {\bibfield  {journal} {\bibinfo
  {journal} {J. Theor. Biol.}\ }\textbf {\bibinfo {volume} {361}},\ \bibinfo
  {pages} {111} (\bibinfo {year} {2014})}\BibitemShut {NoStop}%
\bibitem [{\citenamefont {Chen}\ \emph {et~al.}(2014)\citenamefont {Chen},
  \citenamefont {Szolnoki},\ and\ \citenamefont {Perc}}]{chen2014}%
  \BibitemOpen
  \bibfield  {author} {\bibinfo {author} {\bibfnamefont {X.}~\bibnamefont
  {Chen}}, \bibinfo {author} {\bibfnamefont {A.}~\bibnamefont {Szolnoki}}, \
  and\ \bibinfo {author} {\bibfnamefont {M.}~\bibnamefont {Perc}},\ }\href
  {\doibase 10.1088/1367-2630/16/8/083016} {\bibfield  {journal} {\bibinfo
  {journal} {New J. Phys.}\ }\textbf {\bibinfo {volume} {16}},\ \bibinfo
  {pages} {083016} (\bibinfo {year} {2014})}\BibitemShut {NoStop}%
\bibitem [{\citenamefont {Traulsen}\ \emph {et~al.}(2012)\citenamefont
  {Traulsen}, \citenamefont {R{\"o}hl},\ and\ \citenamefont
  {Milinski}}]{traulsen2012}%
  \BibitemOpen
  \bibfield  {author} {\bibinfo {author} {\bibfnamefont {A.}~\bibnamefont
  {Traulsen}}, \bibinfo {author} {\bibfnamefont {T.}~\bibnamefont {R{\"o}hl}},
  \ and\ \bibinfo {author} {\bibfnamefont {M.}~\bibnamefont {Milinski}},\
  }\href {http://dx.doi.org/10.1098/rspb.2012.0937} {\bibfield  {journal}
  {\bibinfo  {journal} {Proc. R. Soc. B}\ ,\ \bibinfo {pages} {rspb20120937}}
  (\bibinfo {year} {2012})}\BibitemShut {NoStop}%
\bibitem [{\citenamefont {Nikiforakis}(2008)}]{nikiforakis2008}%
  \BibitemOpen
  \bibfield  {author} {\bibinfo {author} {\bibfnamefont {N.}~\bibnamefont
  {Nikiforakis}},\ }\href {\doibase 10.1016/j.jpubeco.2007.04.008} {\bibfield
  {journal} {\bibinfo  {journal} {J. Public Econ.}\ }\textbf {\bibinfo {volume}
  {92}},\ \bibinfo {pages} {91} (\bibinfo {year} {2008})}\BibitemShut {NoStop}%
\bibitem [{\citenamefont {Insko}\ \emph {et~al.}(1985)\citenamefont {Insko},
  \citenamefont {Smith}, \citenamefont {Alicke}, \citenamefont {Wade},\ and\
  \citenamefont {Taylor}}]{insko1985}%
  \BibitemOpen
  \bibfield  {author} {\bibinfo {author} {\bibfnamefont {C.~A.}\ \bibnamefont
  {Insko}}, \bibinfo {author} {\bibfnamefont {R.~H.}\ \bibnamefont {Smith}},
  \bibinfo {author} {\bibfnamefont {M.~D.}\ \bibnamefont {Alicke}}, \bibinfo
  {author} {\bibfnamefont {J.}~\bibnamefont {Wade}}, \ and\ \bibinfo {author}
  {\bibfnamefont {S.}~\bibnamefont {Taylor}},\ }\href {\doibase
  10.1177/0146167285111004} {\bibfield  {journal} {\bibinfo  {journal} {Pers.
  Soc. Psychol. Bull.}\ }\textbf {\bibinfo {volume} {11}},\ \bibinfo {pages}
  {41} (\bibinfo {year} {1985})}\BibitemShut {NoStop}%
\bibitem [{\citenamefont {Campbell}\ and\ \citenamefont
  {Fairey}(1989)}]{campbell1989}%
  \BibitemOpen
  \bibfield  {author} {\bibinfo {author} {\bibfnamefont {J.~D.}\ \bibnamefont
  {Campbell}}\ and\ \bibinfo {author} {\bibfnamefont {P.~J.}\ \bibnamefont
  {Fairey}},\ }\href {\doibase 10.1037/0022-3514.57.3.457} {\bibfield
  {journal} {\bibinfo  {journal} {J. Pers. Soc. Psychol.}\ }\textbf {\bibinfo
  {volume} {57}},\ \bibinfo {pages} {457} (\bibinfo {year} {1989})}\BibitemShut
  {NoStop}%
\bibitem [{\citenamefont {Coleman}(2004)}]{coleman2004}%
  \BibitemOpen
  \bibfield  {author} {\bibinfo {author} {\bibfnamefont {S.}~\bibnamefont
  {Coleman}},\ }\href {\doibase 10.1093/pan/mpg015} {\bibfield  {journal}
  {\bibinfo  {journal} {Political Anal.}\ }\textbf {\bibinfo {volume} {12}},\
  \bibinfo {pages} {76} (\bibinfo {year} {2004})}\BibitemShut {NoStop}%
\bibitem [{\citenamefont {Yang}\ \emph {et~al.}(2013)\citenamefont {Yang},
  \citenamefont {Zhang},\ and\ \citenamefont {Zhou}}]{yang2013}%
  \BibitemOpen
  \bibfield  {author} {\bibinfo {author} {\bibfnamefont {Z.}~\bibnamefont
  {Yang}}, \bibinfo {author} {\bibfnamefont {Z.-K.}\ \bibnamefont {Zhang}}, \
  and\ \bibinfo {author} {\bibfnamefont {T.}~\bibnamefont {Zhou}},\ }\href
  {http://dx.doi.org/10.1093/pan/mpg015} {\bibfield  {journal} {\bibinfo
  {journal} {Europhys. Lett.}\ }\textbf {\bibinfo {volume} {100}},\ \bibinfo
  {pages} {68002} (\bibinfo {year} {2013})}\BibitemShut {NoStop}%
\bibitem [{\citenamefont {Cui}\ and\ \citenamefont {Wu}(2013)}]{cui2013}%
  \BibitemOpen
  \bibfield  {author} {\bibinfo {author} {\bibfnamefont {P.-B.}\ \bibnamefont
  {Cui}}\ and\ \bibinfo {author} {\bibfnamefont {Z.-X.}\ \bibnamefont {Wu}},\
  }\href {\doibase 10.1016/j.physa.2012.10.039} {\bibfield  {journal} {\bibinfo
   {journal} {Physica A}\ }\textbf {\bibinfo {volume} {392}},\ \bibinfo {pages}
  {1500} (\bibinfo {year} {2013})}\BibitemShut {NoStop}%
\bibitem [{\citenamefont {Szolnoki}\ and\ \citenamefont
  {Perc}(2016)}]{szolnoki2016}%
  \BibitemOpen
  \bibfield  {author} {\bibinfo {author} {\bibfnamefont {A.}~\bibnamefont
  {Szolnoki}}\ and\ \bibinfo {author} {\bibfnamefont {M.}~\bibnamefont
  {Perc}},\ }\href {\doibase 10.1038/srep23633} {\bibfield  {journal} {\bibinfo
   {journal} {Sci. Rep.}\ }\textbf {\bibinfo {volume} {6}},\ \bibinfo {pages}
  {23633} (\bibinfo {year} {2016})}\BibitemShut {NoStop}%
\bibitem [{\citenamefont {Hilbe}\ \emph {et~al.}(2014)\citenamefont {Hilbe},
  \citenamefont {Traulsen}, \citenamefont {R{\"o}hl},\ and\ \citenamefont
  {Milinski}}]{hilbe2014}%
  \BibitemOpen
  \bibfield  {author} {\bibinfo {author} {\bibfnamefont {C.}~\bibnamefont
  {Hilbe}}, \bibinfo {author} {\bibfnamefont {A.}~\bibnamefont {Traulsen}},
  \bibinfo {author} {\bibfnamefont {T.}~\bibnamefont {R{\"o}hl}}, \ and\
  \bibinfo {author} {\bibfnamefont {M.}~\bibnamefont {Milinski}},\ }\href
  {\doibase 10.1073/pnas.1315273111} {\bibfield  {journal} {\bibinfo  {journal}
  {Proc. Natl. Acad. Sci. U. S. A.}\ }\textbf {\bibinfo {volume} {111}},\
  \bibinfo {pages} {752} (\bibinfo {year} {2014})}\BibitemShut {NoStop}%
\bibitem [{\citenamefont {Horne}\ and\ \citenamefont
  {Irwin}(2016)}]{horne2016}%
  \BibitemOpen
  \bibfield  {author} {\bibinfo {author} {\bibfnamefont {C.}~\bibnamefont
  {Horne}}\ and\ \bibinfo {author} {\bibfnamefont {K.}~\bibnamefont {Irwin}},\
  }\href {\doibase 10.1080/15534510.2015.1132255} {\bibfield  {journal}
  {\bibinfo  {journal} {Social Influence}\ }\textbf {\bibinfo {volume} {11}},\
  \bibinfo {pages} {7} (\bibinfo {year} {2016})}\BibitemShut {NoStop}%
\bibitem [{\citenamefont {Brown}(2000)}]{brown2000}%
  \BibitemOpen
  \bibfield  {author} {\bibinfo {author} {\bibfnamefont {R.}~\bibnamefont
  {Brown}},\ }\href@noop {} {\emph {\bibinfo {title} {Group processes: Dynamics
  within and between groups (2nd)}}}\ (\bibinfo  {publisher} {Basil
  Blackwell},\ \bibinfo {year} {2000})\BibitemShut {NoStop}%
\bibitem [{\citenamefont {Watts}\ and\ \citenamefont
  {Strogatz}(1998)}]{watts1998}%
  \BibitemOpen
  \bibfield  {author} {\bibinfo {author} {\bibfnamefont {D.~J.}\ \bibnamefont
  {Watts}}\ and\ \bibinfo {author} {\bibfnamefont {S.~H.}\ \bibnamefont
  {Strogatz}},\ }\href {\doibase 10.1038/30918} {\bibfield  {journal} {\bibinfo
   {journal} {Nature}\ }\textbf {\bibinfo {volume} {393}},\ \bibinfo {pages}
  {440} (\bibinfo {year} {1998})}\BibitemShut {NoStop}%
\bibitem [{\citenamefont {Newman}\ and\ \citenamefont
  {Watts}(1999)}]{newman1999}%
  \BibitemOpen
  \bibfield  {author} {\bibinfo {author} {\bibfnamefont {M.~E.}\ \bibnamefont
  {Newman}}\ and\ \bibinfo {author} {\bibfnamefont {D.~J.}\ \bibnamefont
  {Watts}},\ }\href {\doibase 10.1103/PhysRevE.60.7332} {\bibfield  {journal}
  {\bibinfo  {journal} {Phys. Rev. E}\ }\textbf {\bibinfo {volume} {60}},\
  \bibinfo {pages} {7332} (\bibinfo {year} {1999})}\BibitemShut {NoStop}%
\bibitem [{\citenamefont {Szab{\'o}}\ and\ \citenamefont
  {Fath}(2007)}]{szabo2007}%
  \BibitemOpen
  \bibfield  {author} {\bibinfo {author} {\bibfnamefont {G.}~\bibnamefont
  {Szab{\'o}}}\ and\ \bibinfo {author} {\bibfnamefont {G.}~\bibnamefont
  {Fath}},\ }\href {\doibase 10.1016/j.physrep.2007.04.004} {\bibfield
  {journal} {\bibinfo  {journal} {Phys. Rep.}\ }\textbf {\bibinfo {volume}
  {446}},\ \bibinfo {pages} {97} (\bibinfo {year} {2007})}\BibitemShut
  {NoStop}%
\bibitem [{\citenamefont {Szolnoki}\ \emph {et~al.}(2009)\citenamefont
  {Szolnoki}, \citenamefont {Perc},\ and\ \citenamefont
  {Szab{\'o}}}]{szolnoki2009}%
  \BibitemOpen
  \bibfield  {author} {\bibinfo {author} {\bibfnamefont {A.}~\bibnamefont
  {Szolnoki}}, \bibinfo {author} {\bibfnamefont {M.}~\bibnamefont {Perc}}, \
  and\ \bibinfo {author} {\bibfnamefont {G.}~\bibnamefont {Szab{\'o}}},\ }\href
  {\doibase 10.1103/PhysRevE.80.056109} {\bibfield  {journal} {\bibinfo
  {journal} {Phys. Rev. E}\ }\textbf {\bibinfo {volume} {80}},\ \bibinfo
  {pages} {056109} (\bibinfo {year} {2009})}\BibitemShut {NoStop}%
\bibitem [{\citenamefont {Krapivsky}\ and\ \citenamefont
  {Redner}(2003)}]{krapivsky2003}%
  \BibitemOpen
  \bibfield  {author} {\bibinfo {author} {\bibfnamefont {P.~L.}\ \bibnamefont
  {Krapivsky}}\ and\ \bibinfo {author} {\bibfnamefont {S.}~\bibnamefont
  {Redner}},\ }\href {\doibase 10.1103/PhysRevLett.90.238701} {\bibfield
  {journal} {\bibinfo  {journal} {Phys. Rev. Lett.}\ }\textbf {\bibinfo
  {volume} {90}},\ \bibinfo {pages} {238701} (\bibinfo {year}
  {2003})}\BibitemShut {NoStop}%
\bibitem [{\citenamefont {Brandt}\ \emph {et~al.}(2003)\citenamefont {Brandt},
  \citenamefont {Hauert},\ and\ \citenamefont {Sigmund}}]{brandt2003}%
  \BibitemOpen
  \bibfield  {author} {\bibinfo {author} {\bibfnamefont {H.}~\bibnamefont
  {Brandt}}, \bibinfo {author} {\bibfnamefont {C.}~\bibnamefont {Hauert}}, \
  and\ \bibinfo {author} {\bibfnamefont {K.}~\bibnamefont {Sigmund}},\ }\href
  {\doibase 10.1098/rspb.2003.2336} {\bibfield  {journal} {\bibinfo  {journal}
  {Proc. R. Soc. London Ser. B}\ }\textbf {\bibinfo {volume} {270}},\ \bibinfo
  {pages} {1099} (\bibinfo {year} {2003})}\BibitemShut {NoStop}%
\bibitem [{\citenamefont {Fu}\ \emph {et~al.}(2007)\citenamefont {Fu},
  \citenamefont {Liu},\ and\ \citenamefont {Wang}}]{fu2007}%
  \BibitemOpen
  \bibfield  {author} {\bibinfo {author} {\bibfnamefont {F.}~\bibnamefont
  {Fu}}, \bibinfo {author} {\bibfnamefont {L.-H.}\ \bibnamefont {Liu}}, \ and\
  \bibinfo {author} {\bibfnamefont {L.}~\bibnamefont {Wang}},\ }\href
  {https://doi.org/10.1140/epjb/e2007-00124-5} {\bibfield  {journal} {\bibinfo
  {journal} {Eur. Phys. J. B}\ }\textbf {\bibinfo {volume} {56}},\ \bibinfo
  {pages} {367} (\bibinfo {year} {2007})}\BibitemShut {NoStop}%
\bibitem [{\citenamefont {Santos}\ \emph {et~al.}(2008)\citenamefont {Santos},
  \citenamefont {Santos},\ and\ \citenamefont {Pacheco}}]{santos2008}%
  \BibitemOpen
  \bibfield  {author} {\bibinfo {author} {\bibfnamefont {F.~C.}\ \bibnamefont
  {Santos}}, \bibinfo {author} {\bibfnamefont {M.~D.}\ \bibnamefont {Santos}},
  \ and\ \bibinfo {author} {\bibfnamefont {J.~M.}\ \bibnamefont {Pacheco}},\
  }\href {\doibase 10.1038/nature06940} {\bibfield  {journal} {\bibinfo
  {journal} {Nature}\ }\textbf {\bibinfo {volume} {454}},\ \bibinfo {pages}
  {213} (\bibinfo {year} {2008})}\BibitemShut {NoStop}%
\bibitem [{\citenamefont {Szolnoki}\ and\ \citenamefont
  {Perc}(2017)}]{szolnoki2017}%
  \BibitemOpen
  \bibfield  {author} {\bibinfo {author} {\bibfnamefont {A.}~\bibnamefont
  {Szolnoki}}\ and\ \bibinfo {author} {\bibfnamefont {M.}~\bibnamefont
  {Perc}},\ }\href {\doibase 10.1103/PhysRevX.7.041027} {\bibfield  {journal}
  {\bibinfo  {journal} {Phys. Rev. X}\ }\textbf {\bibinfo {volume} {7}},\
  \bibinfo {pages} {041027} (\bibinfo {year} {2017})}\BibitemShut {NoStop}%
\bibitem [{\citenamefont {Weibull}(1997)}]{weibull1997}%
  \BibitemOpen
  \bibfield  {author} {\bibinfo {author} {\bibfnamefont {J.~W.}\ \bibnamefont
  {Weibull}},\ }\href@noop {} {\emph {\bibinfo {title} {Evolutionary game
  theory}}}\ (\bibinfo  {publisher} {MIT press},\ \bibinfo {year}
  {1997})\BibitemShut {NoStop}%
\end{thebibliography}%

\end{document}